\documentclass[twocolumn,floatfix,prl]{revtex4}%
\usepackage[dvipdfmx]{graphicx}
\usepackage[dvipdfmx]{color}
\usepackage{amsmath,amssymb,amsfonts}
\usepackage{mathrsfs}
\usepackage{mathtools}
\usepackage{here}
\usepackage{bm}
\usepackage[dvipdfmx]{color}

\def\s{{\sigma}}
\def\e{{\epsilon}}
\def\k{{ {\bm k} }}

\def\q{{ {\bm q} }}

\def\0{{ {\bm 0} }}

\def\a{{\alpha}}

\def\r{{ {\bm r} }}

\allowdisplaybreaks[4]

\begin{document}
\title{
Origin of switchable quasiparticle-interference chirality in loop-current phase of kagome metals measured by scanning-tunneling-microscopy 
}
\author{
Seigo Nakazawa$^1$, Rina Tazai$^2$, Youichi Yamakawa$^1$, Seiichiro Onari$^1$, and Hiroshi Kontani$^1$
}
\date{\today } 

\begin{abstract}
The chiral loop-current (LC) phase in kagome metals AV$_3$Sb$_5$ (A = Cs, Rb, K) has attracted considerable attention as a novel quantum state driven by electron correlations.
Scanning tunneling microscopy (STM) experiments have provided strong evidence for the chiral LC phase through the detection of chirality in the quasiparticle interference (QPI) signal.
However, the fundamental relationship between ``QPI chirality'' and ``LC chirality'' remains unexplored. 
For instance, the QPI signal is unchanged even when all LC orders are inverted. 
Furthermore, only the chiral LC order cannot induce QPI chirality. 
At present, the true essence of kagome metals that we should learn from the remarkable QPI experiments remains elusive.
To address this, we investigate the origin of the QPI signal in the LC phase using a large unit-cell tight-binding model for kagome metals.
The LC phase gives rise to a $Z_3$ nematic phase, characterized by three distinct directors, under the Star-of-David bond order.
Our findings demonstrate that the QPI chirality induced by a single impurity at site Z, denoted as $\chi_{\rm Z}$, can take values of $\pm1$ (chiral) or $0$ (achiral), depending on the direction of the $Z_3$ nematic order.
Prominent QPI chirality originates from extremely dilute impurities ($\lesssim0.1$ \%) in the present mechanism. 
Notably, $\chi_{\rm Z}$ ($=\pm1$, $0$) changes smoothly with minimal free-energy barriers by applying a small magnetic field $B_z$, accompanied by a switching of the $Z_3$ nematic director.
This study provides a comprehensive explanation for the observed ``$B_z$-switchable QPI chirality'' in regions with dilute impurities, offering fundamental insight into the chiral LC in kagome metals.



\end{abstract}

\affiliation{
$^1$Department of Physics, Nagoya University,
Furo-cho, Nagoya 464-8602, Japan \\
$^2$Yukawa Institute for Theoretical Physics, Kyoto University, Kyoto 606-8502, Japan.
}
\sloppy

\maketitle

\section*{Introduction}
In strongly correlated metals, a variety of quantum phase transitions have been discovered over the years, leading to intriguing electronic states due to the breaking of symmetries like time-reversal symmetry (TRS), rotational symmetry, and chirality. 
The recently discovered kagome metal AV$_3$Sb$_5$ (A = Cs, Rb, K) \cite{Ortiz1,Ortiz2} displays remarkable electronic states arising from a sequence of quantum phases, including charge density wave (CDW) \cite{BO2,BO3,BO4}, nematic states \cite{BO2,nematicity1,nematicity2,torque,Tazai1}, and superconductivity \cite{SC1,SC2}. 
Particularly, the TRS-breaking loop-current (LC) state, which lacks spin polarization, has garnered significant attention \cite{STM,STM-Yin,STM-Mad,STM-PRB1,STM-PRB2}. 
The LC order parameter involves a ``pure imaginary'' modulation in the hopping integral, $\delta t^{\rm c} = \pm i\eta$ 
\cite{Varma,Varma2,Balents,cLC,BO_theory1,BO_theory2,BO_theory3,BO_theory4,Tazai1,Fernandes}.

The kagome lattice structure and the Fermi surfaces without density waves are shown in Figure \ref{fig:fig1} (a).
Figure \ref{fig:fig1} (b) depicts the ``chiral LC'' state that violates any 2D mirror symmetries, characterized by $\delta t^{\rm c} \ne 0$ in all ab-, bc-, and ca-directions \cite{Balents,cLC,BO_theory1,BO_theory2,BO_theory3,BO_theory4,Tazai1,Fernandes}. 
This state is associated with a finite uniform orbital magnetization ($M_{\rm orb}$), which can be switched by a small out-of-plane magnetic field $B_z$, as evidenced by the anomalous Hall effect (AHE) \cite{AHE1,AHE2,AHE3} and electronic magneto-chiral anisotropy (eMChA) \cite{eMChA}. 
The CDW order parameter, $\delta t^{\rm b} = \pm \phi$ (real), is also illustrated in Fig. \ref{fig:fig1} (b) \cite{Tazai1};
see Supplementary Information (SI) A for details \cite{SM}.
The microscopic mechanism of the LC order has been investigated in Refs.
\cite{Balents,cLC,BO_theory1,BO_theory2,BO_theory3,BO_theory4,Tazai1,Fernandes},
including the BO fluctuation-mediated mechanism \cite{Tazai2}.


Experimentally, the TRS breaking electronic states in the chiral LC phase
have been reported by scanning tunneling microscopy (STM) \cite{STM,STM-Yin,STM-Mad,STM-PRB1,STM-PRB2},
$\mu$SR \cite{mSR1,mSR2,mSR3,mSR4}, NMR \cite{NMR}, Kerr effect \cite{nematicity2,Kerr2}, and AHE \cite{AHE1,AHE2,AHE3} measurements.
Notably, nonreciprocal transport measurement by eMChA \cite{eMChA} 
reveals the violations of both TRS and inversion symmetries.
Many of these experiments indicate the chiral LC appears at $\sim T_{\rm lc}$ inside the CDW phase ($T<T_{\rm cdw}\sim100$K).
Recent study \cite{strain} shows that $T_{\rm lc}$ drastically increases under the weak magnetic field $B_z$ or tiny strain $\e$,
which is naturally explained based on the extended Ginzburg-Landau free-energy theory \cite{Tazai3}. 
Additionally, magnetic torque measurements \cite{torque} suggest that an achiral LC without $M_{\rm orb}$ appears around $\sim 130$ K.
Theoretically, achiral to chiral LC transition is triggered by the uniform magnetic field or the CDW order parameter below $T_{\rm cdw}$.

Intuitive and strong evidence of the chiral LC order below $T_{\rm cdw}$ comes from the STM measurements, which reveal chirality in the quasiparticle interference (QPI) signal induced by dilute impurities. 
The QPI signal is the Fourier transform of the local density of states (LDOS) over a large area, including $1,000-10,000$ sites.
(QPI experiments can reveal fundamental quantum orders in metals, including the nematic/smectic density-wave states \cite{STM-Fujita,STM-Hanaguri} and the pair-density-wave states \cite{STM-PDW}.)
Notably, the observed chirality in the QPI signal in the LC phase is switched by small $B_z$ in A=K \cite{STM,STM-Yin}, A=Cs \cite{STM-Yin,STM-PRB1}, and A=Rb \cite{STM-Mad,STM-PRB2} compounds.
However, the fundamental relationship between QPI chirality and LC chirality remains unexplored. 
For instance, the QPI signal remains unchanged even when all LC orders are inverted (by $B_z$), because the LDOS is TRS-even ({\it i.e.}, $\eta$-even) and thus unaffected. 
Furthermore, the presence of only the chiral LC order cannot induce QPI chirality, despite the induction of a finite $M_{\rm orb}$. 
Consequently, despite the significance of QPI experiments in kagome metals \cite{STM,STM-Yin,STM-PRB1,STM-Mad,STM-PRB2}, the underlying origin of QPI chirality remains poorly understood.

To address this unresolved issue, in this paper,
we investigate the origin of the QPI signal in the LC phase using a large unit-cell model for kagome metals. 
The coexistence of LC and BO gives rise to the three-directional ($Z_3$) nematic phase.
Our findings show that the QPI chirality due to the single impurity at site Z, denoted as $\chi_{\rm Z}$, takes $\pm1$ (chiral) or $0$ (achiral), depending on the director of the $Z_3$ LC + BO nematic order. 
Importantly, prominent QPI chirality originates from extremely dilute impurities ($\lesssim0.1$\%).
Remarkably, $\chi_{\rm Z}$ changes by applying a small magnetic field $B_z$, accompanied by a change of $Z_3$ nematic director with minimal free-energy barriers.
This study provides a highly promising explanation for the $B_z$-reversible QPI chirality observed in regions with dilute impurities, offering essential insight into the chiral LC phase in kagome metals.


The microscopic mechanism of the exotic multiple quantum phase transitions has been studied very actively. 
The mean-field theories as well as the renormalization group theories based on the (extended) Hubbard models have been performed in Refs. 
\cite{BO_theory3,BO_theory4,Balents,Tazai1,Tazai2,Thomale,Sushkov,Shimura}.
We discovered that the Star-of-David (SoD) BO state is driven by the ``paramagnon-interference'' mechanism \cite{Tazai1}, 
which is described by the Aslamazov-Larkin vertex corrections that are dropped in the mean-field level approximations
\cite{Onari,fRG,DW_equation,Jianxin,Kontani-rev}.
Importantly, the BO fluctuations mediate not only $s$-wave or $p$-wave superconductivity but also the TRS breaking LC order
\cite{Tazai1,Tazai2}.


\section*{Kagome lattice model with LC + BO orders}
\begin{figure}[htb]
\includegraphics[width=0.99\linewidth]{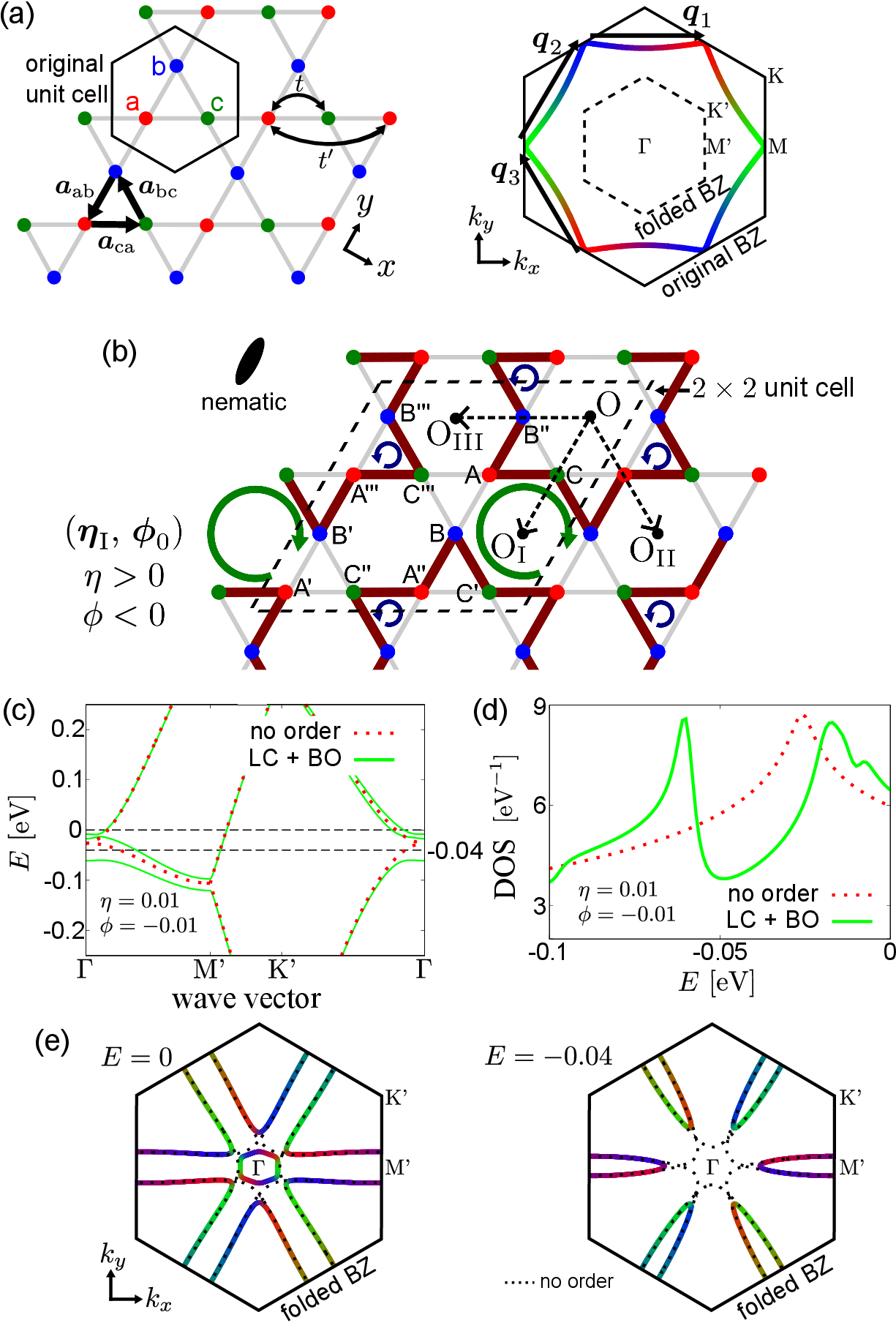}
\caption{
(a) Kagome lattice model.
The unit cell is composed of the sublattices a, b, and c.
$2\boldsymbol{a}_{\mathrm{ab}}$, $2\boldsymbol{a}_{\mathrm{bc}}$
give the two primitive vectors.
Fermi surfaces without density waves are also shown.
Red, blue, and green colors represent the weight of the sublattices a, b, and c, respectively.
(b) Nematic LC + BO existing state with ${\bm\phi}_0$ and ${\bm\eta}_{\rm I}$.
The director is along $y$ (or equivalently $x$) axis. 
The $2\times2$ unit cell is composed of 12 sites ${\rm Z}$, ${\rm Z}'$, ${\rm Z}''$, and ${\rm Z}'''$ with Z = A, B, C.
The $C_6$ center of ${\bm\eta}_\a$ is denoted as $O_\a$. 
(c) Band structure and (d) DOS in the folded BZ for $-\phi=\eta=0.01$. 
(e) Equal-energy surface for $E=0$ (Fermi level) and $E=-0.04$.
}
\label{fig:fig1}
\end{figure}

First, we introduce the kagome lattice model \cite{Tazai1,Thomale}
shown in Fig. \ref{fig:fig1} (a).
The original unit cell is composed of three sublattices a, b, and c.
We put the nearest and the next-neighbor hopping integrals, 
$t=-0.5\,\mathrm{eV}$ and $t'=-0.04\,\mathrm{eV}$,
to approximate the real Fermi surface 
and set the temperature $T=0.0025\,\mathrm{eV}$.
Hereafter, the unit of energy is eV.
Fermi surfaces without density waves are also shown.
Here, red, blue, and green colors denote the weight of the sublattices a, b, and c, respectively.
Figure \ref{fig:fig1} (b) shows the enlarged 12 site unit cell
under the triple-$\q$ order.
We introduce the LC order ${\bm\eta}\equiv(\eta_1,\eta_2,\eta_3)$
and the BO ${\bm\phi}\equiv(\phi_1,\phi_2,\phi_3)$,
where $\eta_n$ and $\phi_n$ are the order parameters with the wavevector $\q_n$.
The definitions of the LC ${\bm\eta}$ and the BO ${\bm\phi}$ are given in the SI A.

When $T_{\rm cdw}> T_{\rm lc}$, we can set the BO as
${\bm\phi}_0=\phi{\bm e}_0$ without the loss of generality,
where ${\bm e}_0=(1,1,1)/\sqrt{3}$.
Then, just below $T_{\rm lc}$, the LC order that minimize the GL free energy
is given as ${\bm\eta}_{\a}=\eta{\bm e}_{\a}$ ($\a$=I, II, III),
where ${\bm e}_{\rm I}=(2,-1,-1)/\sqrt{6}$, 
${\bm e}_{\rm II}=(-1,2,-1)/\sqrt{6}$, 
and ${\bm e}_{\rm III}=(-1,-1,2)/\sqrt{6}$.
Figure \ref{fig:fig1} (b) shows the SoD BO by ${\bm\phi}_0$ ($\phi<0$) and the LC order by ${\bm\eta}_{\rm I}$ ($\eta>0$):
The band structure and the density-of-states (DOS) are shown in Figs. \ref{fig:fig1} (c) and (d) in the folded Brillouin zone (BZ) for $\eta=-\phi=0.01$.
The equal-energy surfaces for $E=0$ (Fermi level) and $-0.04$ are shown in Fig. \ref{fig:fig1} (e).

Importantly, the coexisting state $({\bm\eta}_{\a}, \ {\bm\phi}_0)$ is nematic because the $C_6$ symmetry center of the LC, $O_\a$, and that of the BO, $O$, are different.
The director is along $O$-$O_\a$ line;
See the SI A for details \cite{SM}.
In the LC + BO state in kagome metals, dilute impurities cause nontrivial sizable change in the electronic states \cite{Nakazawa-imp}. 
To reveal the origin of the impurity-induced chiral QPI signal in kagome metals, we investigate the single impurity effect based on the giant unit-cell kagome lattice model ($N\sim1500$):
\begin{eqnarray}
H=\sum_\k \sum_{i,j,\s}t_{\k,i,j}c_{\k,i,\s}^\dagger c_{\k,j,\s} + V \sum_\s c_{\k,{\rm Z},\s}^\dagger c_{\k,{\rm Z},\s} ,
\label{eqn:Ham}
\end{eqnarray}
where $i,\ j \ (=0\sim N-1)$ represents the site indices in the unit-cell, and $V$ is the impurity potential at site Z.
Hereafter, we set $V=\infty$.
$\k$ is the wavevector due to the periodic configuration of the giant unit-cell.
The hopping integral is
$t_{\k,i,j}=t^0_{\k,i,j}+\delta t^{\rm c}_{\k,i,j}+\delta t^{\rm b}_{\k,i,j}$,
where $t^0_{\k,i,j}$ is the original hopping integral with $D_{6h}$ symmetry,
$\delta t^{\rm c}_{\k,i,j}$ and $\delta t^{\rm b}_{\k,i,j}$ are the LC and BO parameters; see the SI A  \cite{SM}.
The LDOS at site $i$ is
%
%
$\rho_{i}(E)=\frac1{N_k}\sum_\k \delta(\e_{b,\k}-E) |U_{i,b}(\k)|^2$,
where $b=1\sim N$ is the band index.
Here, $\e_{b,\k}(\k)$ is the Bloch band-energy measured from the chemical potential,  
and $U_{i,b}(\k)$ is the matrix element of the unitary transformation from $i$ to $b$.
We use $\ge 2048^2$ $\k$-meshes in the numerical study.

\section*{QPI signal chirality}
\begin{figure}[htb]
\includegraphics[width=0.99\linewidth]{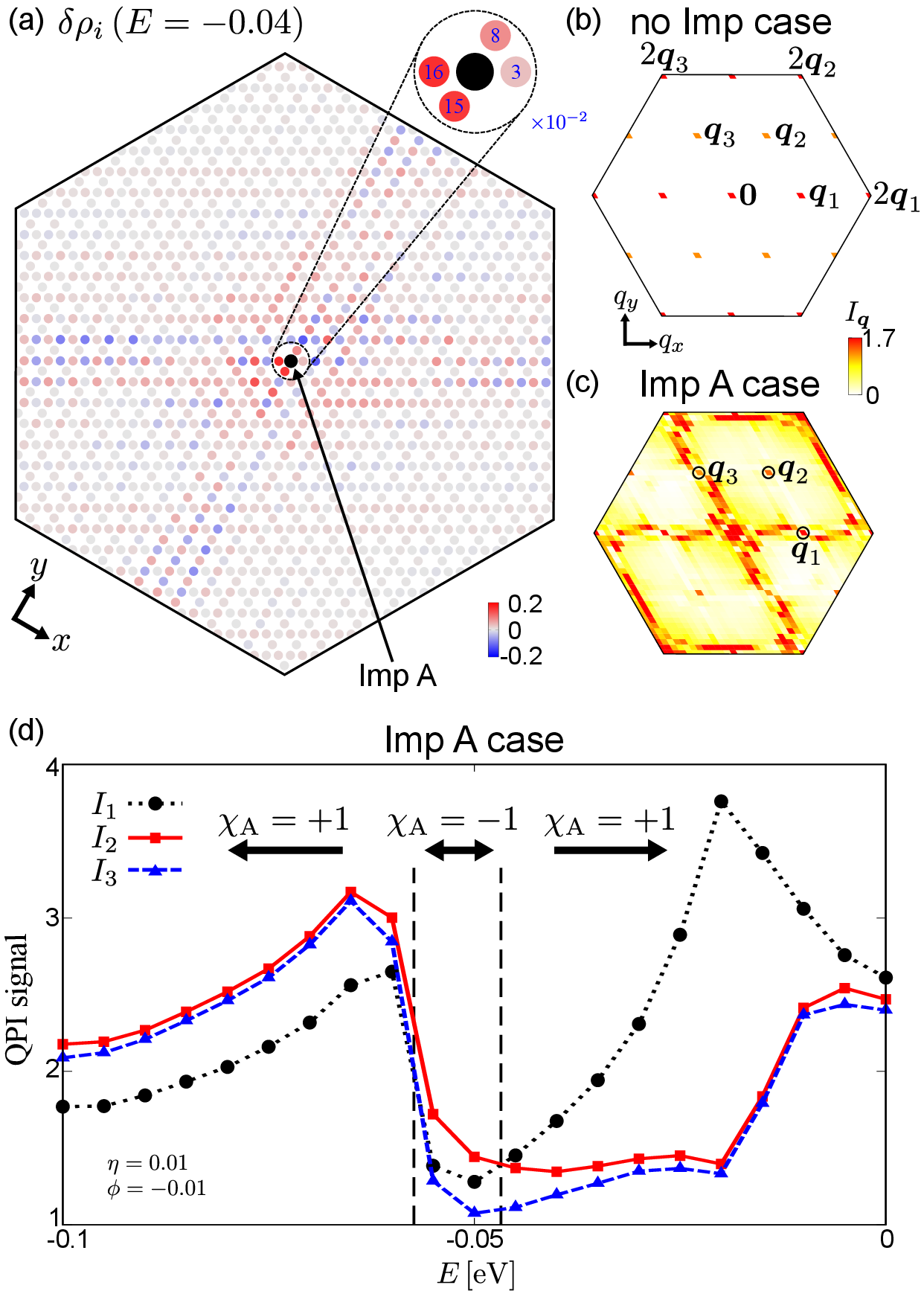}
\caption{
(a) Modulation of the LDOS $\delta\rho_i$ ($E=-0.04$) induced by Imp A in $N=1452$-site unit cell under the LC + BO state with (${\bm\eta}_{\rm I}$, ${\bm\phi}_0$). 
Single impurity gives drastic long-range modulation, reflecting the chirality of the LC phase.
(Inset) LDOS of the four sites  around Imp A.
(b) QPI signals $I_\q$ for $V=0$.
$2\q_n$ is the Bragg peak of the 3-site unit-cell, and $\q_n$ is the Bragg peak of the 12-site due to LC + BO order.
(c) QPI signals for $V=\infty$ given by the Fourier transformation of the LDOS shown in (a).
(d) $E$-dependence of the QPI signal ($I_1,I_2,I_3$) due to a single Imp A, whose chirality is defined as $\chi_{\rm A}\equiv\varepsilon_{n_1 n_2 n_3}$ for $I_{n_1}>I_{n_2}>I_{n_3}$. 
}
\label{fig:fig2}
\end{figure}

Figure \ref{fig:fig2} (a) represents 
the modulation of the LDOS ($\delta\rho_i\equiv\rho_i-\rho_i^{V=0}$) at $E=-0.04$ induced by the single impurity at A-site (=Imp A), whose location is shown in Fig. \ref{fig:fig1} (b).
We set $N=12M_x M_y$-site unit cell with $M_x=M_y=11$ ($N=1452$) under the LC + BO state with $-\phi=\eta=0.01$. 
(The QPI measurements in Refs. \cite{STM,STM-Yin,STM-Mad,STM-PRB1,STM-PRB2} were performed for $N=5000\sim10000$ V sites.)
Thus, a single impurity induces long-range modulation in $\rho_i(E)$, driven by significant Friedel oscillations in the LC + BO phase \cite{Nakazawa-imp}.
The present LDOS pattern is achiral because any 2D mirror symmetries are violated.
This fact leads to the chiral QPI signal, as we will discuss in the paper.
(The LDOS pattern without LC exhibit mirror symmetry; see the SI A \cite{SM}.) 
The symmetry of the obtained LDOS pattern governs the QPI signals, which is given by the Fourier transformation of the LDOS:
%
\begin{eqnarray}
I_\q(E) &=& \left|\sum_{i}\rho_i(E)e^{i\q\cdot\r_i}\right| 
\nonumber \\
&=& \left[ \sum_{i,j}\rho_i(E)\rho_j(E) \cos(\q\cdot(\r_i-\r_j)) \right]^{\frac12}.
\end{eqnarray}
Without impurities, the QPI signal exhibits sharp Bragg peaks at $\q=2\q_n$ and $\q=\q_n$, as shown in Fig. \ref{fig:fig2} (b).
The former (latter) originates from the 3-site (the 12-site) unit-cell.
The QPI signal exhibits the nematic (achiral) peak structure; $I_1\ne I_2 = I_3$ ($I_n\equiv I_{\q_n}$).
Notably, a single impurity causes a dramatic change in the QPI signal in the large site model, as shown in Fig. \ref{fig:fig2} (c).
Figure \ref{fig:fig2} (d) exhibits the $E$-dependence of the QPI signal due to Imp A for $N=1452$ model.
Importantly, the QPI signal exhibits the chiral peak structure; $I_1\ne I_2 \ne I_3$.
Here, we define the QPI chirality due to Imp A as $\chi_{\rm A}\equiv\varepsilon_{n_1 n_2 n_3}$ for $I_{n_1}>I_{n_2}>I_{n_3}$. 
($\varepsilon_{n m l}$ is a Levi-Civita tensor.)
Interestingly, $\chi_{\rm A}$ depends on the energy $E$.
These results are consistent with experimental reports \cite{STM,STM-Yin,STM-Mad,STM-PRB1,STM-PRB2}.
Thus, the chiral QPI signal in kagome metals is naturally explained by the LC + BO nematic state predicted in Ref. \cite{Tazai2}. 
This is the main finding of this study.


\begin{figure}[htb]
\includegraphics[width=0.99\linewidth]{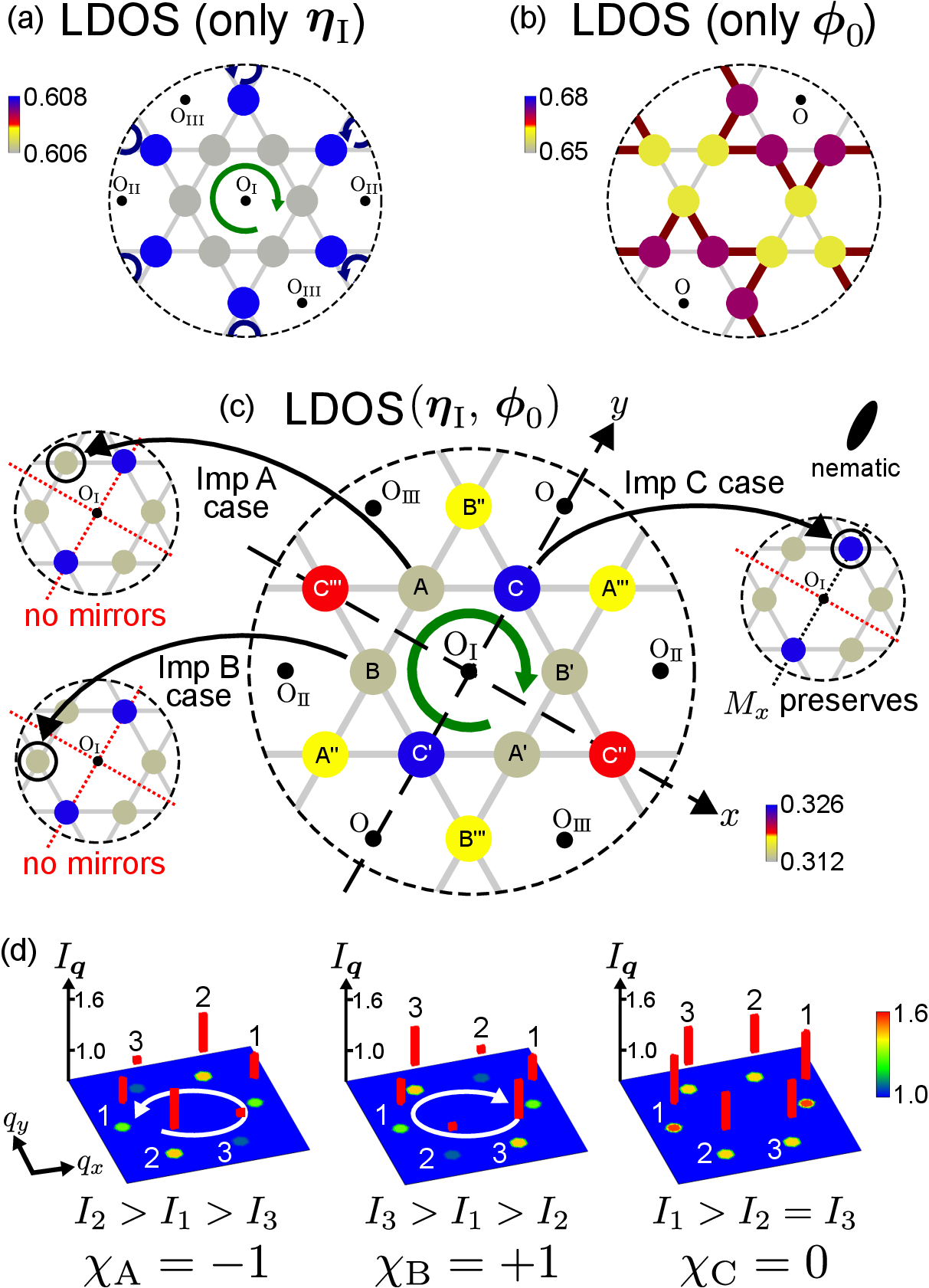}
\caption{
(a) LDOS under the LC order $(\eta,\phi)=(0.01,0)$ and (b) LDOS under the BO order $(\eta,\phi)=(0,-0.01)$ for $V=0$ at $E=-0.05$. 
Both states have $C_6$ symmetry.
(c) LDOS under the LC + BO order [$\eta=-\phi=0.01$] with $C_2$ symmetry.
All mirror symmetries are violated by a single impurity at A-site or B-site.
Therefore, the chirality is induced by a single impurity.
In contrast, a mirror symmetries survive by the C-site impurity.
(d) Obtained QPI signal at $E=-0.05$, which is chiral for A-site impurity ($\chi_{\rm A}=-1$) and B-site impurity ($\chi_{\rm B}=+1$), while it is achiral for C-site impurity ($\chi_{\rm C}=0$).
}
\label{fig:fig3}
\end{figure}

Here, we discuss the origin of the QPI chirality from the perspective of symmetry.
Figure \ref{fig:fig3} (a) [(b)] shows the LDOS in the LC [BO] state at $E=-0.05$ for $V=0$.
In both LC and BO states, $C_6$ symmetry is preserved for around $O_{\rm I}$ and $O$, respectively.
Because $O_{\rm I}\ne O$,
the symmetry of the LDOS in the LC + BO state shown in Fig. \ref{fig:fig3} (c) is reduced to $C_2$, while two mirror symmetries $M_x$ ($x\rightarrow-x$) and $M_y$ ($y\rightarrow-y$) survive.
Thus, the LDOS for $V=0$ exhibits four different values.
Now, we discuss the chiral symmetry induced by a single impurity.
For Imp A state, all in-plane mirror symmetries are broken, and it transforms into the Imp B state by $M_y$ operation for $({\bm\eta}_{\rm I},{\bm\phi}_0)$.
Therefore, if $\chi_{\rm A}=-1$ for Imp A state, then $\chi_{\rm B}=+1$ for Imp B state.
In contrast, the Imp C state is unchanged by $M_x$ operation, so the electronic state is achiral $\chi_{\rm C}=0$.
The numerically derived QPI signals for Imp A-C states are depicted in Fig. \ref{fig:fig3} (d).
Thus, when the director of the LC + BO nematicity is along ab direction, the QPI chirality emerges for Imp A and Imp B states, and the relation $\chi_{\rm A}=-\chi_{\rm B}$ holds.
It is noteworthy that the chiral QPI signal appears when two impurities are introduced in sites A and A', where the inversion-symmetry holds. 
Therefore, the chiral QPI originates from the 2D mirror symmetry violations, even if 2D inversion symmetry ($I=M_x M_y$) holds.

\section*{Change of $\chi_{\rm Z}$ by magnetic field}
Now, we explain that the QPI chirality by Imp Z ($\chi_{\rm Z}=\pm1, 0$) is changed by small magnetic field along $z$-axis, $B_z$.
As we explain in the SI B \cite{SM}, 
the GL free energy for $({\bm\eta},{\bm\phi}_0)$ is given as
$F[{\bm\eta},{\bm\phi}_0]= A_c |{\bm\eta}|^2+ B (\eta_1+\eta_2+\eta_3)^2+C|{\bm\eta}|^4+C'(\eta_1^2\eta_2^2+{\rm cycl.})$,
where $\eta_0\equiv|{\bm\eta}|$ and $A_c$, $B$, $C$ and $C'$ are the GL coefficients renormalized by $\phi$.
We assume $\phi$ is independent of $\eta$,
which is justified when $|{\bm\eta}|\ll|\phi|$ for $T_{\rm cdw}\gg T_{\rm lc} \gtrsim T$.
In addition, we can expect the relation $C'\ll C$ and $B>0$ in kagome metals.
Then, $F[{\bm\eta},{\bm\phi}_0]$ is minimized when
(i) $|{\bm\eta}|\equiv\eta_0 \ {\rm (=const.)}$ and (ii) $\eta_1+\eta_2+\eta_3=0$.

\begin{figure}[htb]
\includegraphics[width=0.99\linewidth]{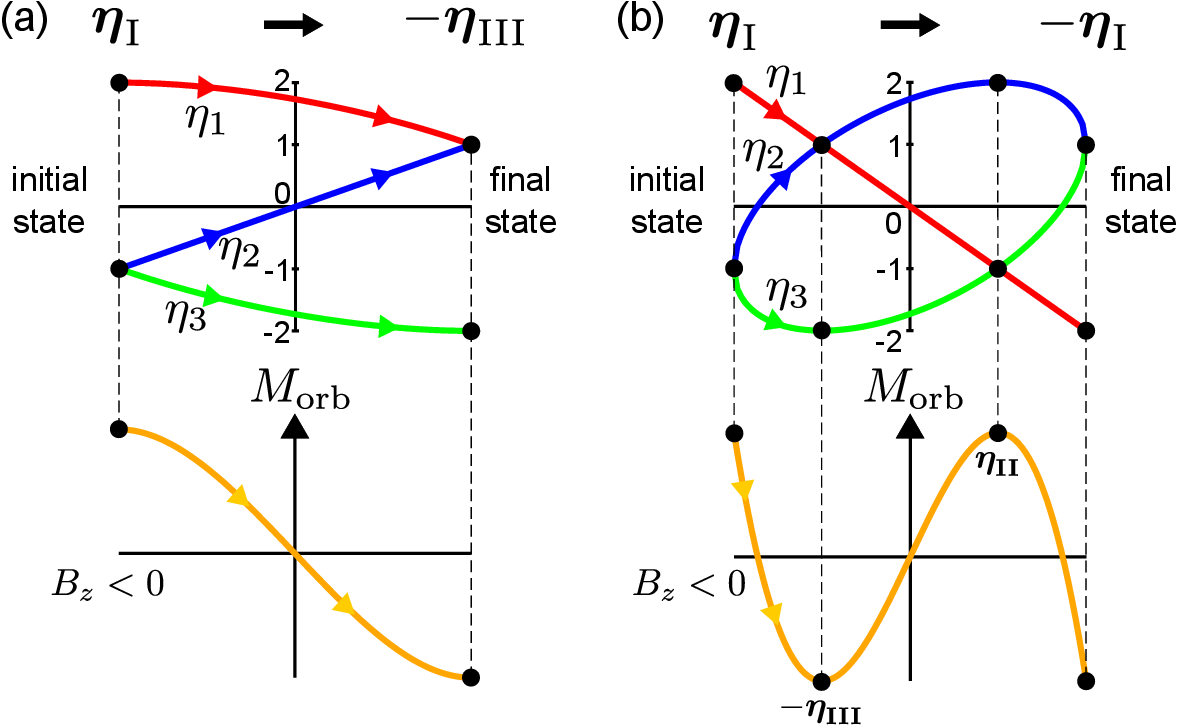}
\caption{
(a) Adiabatic change from ${\bm\eta}_{\rm I}$ to $-{\bm\eta}_{\rm III}$ under the constant $|{\bm\eta}|=\eta_0=\sqrt{6}$ and $\eta_1+\eta_2+\eta_3=0$.
The corresponding $M_{\rm orb}\propto \eta_1 \eta_2 \eta_3$ exhibits a monotonic change.
Therefore, the transition from ${\bm\eta}_{\rm I}$ to $-{\bm\eta}_{\rm III}$ (or $-{\bm\eta}_{\rm II}$) can be induced by the outer magnetic field $B_z$.
(b) Adiabatic change from ${\bm\eta}_{\rm I}$ to $-{\bm\eta}_{\rm I}$ under the constant $|{\bm\eta}|$. 
Because $M_{\rm orb}\propto \eta_1 \eta_2 \eta_3$ is non-monotonic, the transition from ${\bm\eta}_{\rm I}$ to $-{\bm\eta}_{\rm I}$ is not realized.
}
\label{fig:fig4}
\end{figure}

As we explain in the SI B \cite{SM}, the orbital magnetization is $M_{\rm orb}= m_3 \eta_1\eta_2\eta_3$ under the constraint (ii).
Therefore, in the case of $m_3<0$,
the state ${\bm\eta}_{\a}$ is realized under $B_z>0$,
and it is changed to $-{\bm\eta}_{\a'}$ under $B_z<0$.
Now, we consider the adiabatic change from ${\bm\eta}_{\rm I}$ to $-{\bm\eta}_{\a}$ under the constraints (i) and (ii).
Figure \ref{fig:fig4} (a) illustrates the process ${\bm\eta}_{\rm I}\rightarrow -{\bm\eta}_{\rm III}$ as a function of $\eta_{2}$.
This process occurs under $B_z$ because $M_{\rm orb}$ undergoes a monotonic change with a sign reversal.
Importantly, by applying a $+\pi/3$ rotation around the BO center $O$, the state $(-{\bm\eta}_{\rm III},{\bm\phi}_0)$ transforms into $(-{\bm\eta}_{\rm I},{\bm\phi}_0)$, exhibiting the same LDOS pattern shown in Fig. \ref{fig:fig3} (c).
Since the same rotation maps Imp A to Imp B', it follows that $\chi_{\rm A}(-{\bm\eta}_{\rm III},{\bm\phi}_0)$ is equal to $\chi_{\rm B'}(-{\bm\eta}_{\rm I},{\bm\phi}_0)=+1$.
(Note that both the LDOS and the QPI signal remain invariant under the transformation ${\bm\eta}\rightarrow-{\bm\eta}$.)
Similarly, the transition from ${\bm\eta}_{\rm I}$ to $-{\bm\eta}_{\rm II}$ can also be realized, leading to $\chi_{\rm A}(-{\bm\eta}_{\rm II},{\bm\phi}_0)=\chi_{\rm C''}(-{\bm\eta}_{\rm I},{\bm\phi}_0)=0$.
In contrast, the process ${\bm\eta}_{\rm I} \rightarrow -{\bm\eta}_{\rm I}$ shown in Fig. \ref{fig:fig4} (b) would be difficult to be realized under $B_z$ because $M_{\rm orb}$ exhibits non-monotonic behavior.
In summary, the QPI chirality $\chi_{\rm Z} = 0, \pm1$ changes by applying tiny $B_z$. 
This theory provides a consistent explanation for the switchable chiral QPI signal in kagome metals.

The QPI signals due to Imp Z and Imp Z' (Z=A, B, C) shown in Fig. \ref{fig:fig3} (c) are equivalent .
However, the QPI signals due to Imp Z$''$ (Imp Z$'''$) are different from those due to Imp Z, from the perspective of symmetry.
In the SI C \cite{SM}, we show that the magnitude of the QPI signal chirality is very small for Imp Z$''$ case.

\section*{Discussions}
Experimentally, the QPI signal becomes chiral in the dilute impurity region.
When $n_{\rm imp}\approx0.1$\% in $N\approx5000$ site model,
a single impurity is likely to occupy an A, A', B, or B' site in Fig. \ref{fig:fig3} (c), giving rise to the chiral QPI signal.
Furthermore, the Sb-site impurity (or vacancy) also induces sizable chiral QPI, as we discuss in the SI D \cite{SM}.
Note that the QPI signal becomes nematic for impurity-free region for A=K \cite{STM-Yin,BO2}.
These observations are consistently explained by the present theory.
Notably, in the dense impurity region, the QPI signal transitions to a nematic state, possibly due to the suppression of the LC order \cite{Nakazawa-imp}.

In Table I, we summarize the sign of $M_{\rm orb}$, the direction of the nematicity, and the sign of the Imp-Z induced QPI chirality $\chi_{\rm Z}$ in the $Z_3$ nematic LC + BO $Z_3$ state $(\pm{\bm\eta}_\a,{\bm\phi}_0)$.
Here, Z = A, B, or C, and $\a$ = I, II, or III.
In this Table, we define the sign of $M_{\rm orb}$ and $\chi_{\rm A}$ is $+$ for the LC + BO state $({\bm\eta}_{\rm I},{\bm\phi}_0)$.

\begin{table}[htbp]
\caption{$M_{\rm orb}$, director, and Imp-Z induced QPI chirality $\chi_{\rm Z}$ in the $Z_3$ nematic LC + BO state $(\pm{\bm\eta}_\a,{\bm\phi}_0)$.}
\centering
\begin{tabular}{|c|c|c|c|c|c|c|}
\hline
LC & ${\bm\eta}_{\rm I}$ & ${\bm\eta}_{\rm II}$ & ${\bm\eta}_{\rm III}$ & $-{\bm\eta}_{\rm I}$ & $-{\bm\eta}_{\rm II}$ & $-{\bm\eta}_{\rm III}$ 
 \\
\hline
$M_{\rm orb}$ & $+$ & $+$ & $+$ & $-$ & $-$ & $-$  \\ \hline
director & $0$ & $\pi/3$ & $2\pi/3$ & $0$ & $\pi/3$ & $2\pi/3$  \\ \hline
$\chi_{\rm A}$ & $+$ & $0$ & $-$ & $+$ & $0$ & $-$  \\ \hline
$\chi_{\rm B}$ & $-$ & $+$ & $0$ & $-$ & $+$ & $0$  \\ \hline
$\chi_{\rm C}$ & $0$ & $-$ & $+$ & $0$ & $-$ & $+$  \\ \hline
\hline
\end{tabular}
\end{table}

Here, we explore the possibility of impurity-free chiral QPI for A = Cs and Rb, where $2a_0$ and $4a_0$ stripe CDW is commonly observed in STM studies \cite{stripe1, stripe2}.
(Note that this stripe order is absent for A = K.)
As discussed in Ref. \cite{Tazai-eMChA}, the stripe CDW breaks inversion symmetry when it coexists with the LC order.
If the $Z_3$ nematic director is not aligned with the stripe CDW, a chiral QPI signal ($I_1 \ne I_2 \ne I_3$) emerges even in the absence of impurities, as explained in SI E \cite{SM}.

The $3Q$-LC state exhibits weak bulk ferromagnetism ($M_{\rm orb} \ne 0$) when the wavevectors $\q_n$ ($n = 1, 2, 3$) are commensurate.
However, domains with opposite signs of $M_{\rm orb}$ may form if $\q_n$ are slightly incommensurate.
In such cases, applying a small $B_z$ field would restore bulk ferromagnetism 
to gain the Zeeman energy.

In summary, we present the switchable chiral QPI signal in kagome metals as decisive evidence for the emergence of LC order in the BO phase.
Prominent QPI chirality originates from extremely dilute impurities ($\lesssim0.1$\%) in kagome metals.
Interestingly, the chirality of the QPI is found to be strongly correlated with the $Z_3$ nematicity originating from the off-site LC + BO state \cite{Tazai2}.
Furthermore, this study represents an important step toward understanding the chirality in the pair-density-wave state recently reported in kagome metals \cite{STM-Yin}.
In this study, we have omitted the m-type Fermi surface, formed by $d_{yz}$ orbitals, which is quantitatively important for the LC-induced $M_{\rm orb}$ \cite{Tazai3} and eMChA \cite{Tazai-eMChA}.
Therefore, the influence of the m-type Fermi surface on the chiral QPI signal will be an important future issue.




\subsection{Acknowledgments}
We are grateful to Y. Matsuda, T. Shibauchi, K. Hashimoto, T. Asaba, and S. Suetsugu for very useful discussions.









\clearpage

\makeatletter
\renewcommand{\thefigure}{S\arabic{figure}}
\renewcommand{\theequation}{S\arabic{equation}}
\makeatother
\setcounter{figure}{0}
\setcounter{equation}{0}
\setcounter{page}{1}
\setcounter{section}{1}

\begin{widetext}
\begin{center}
{\bf 
[Supplementary Information] \\
Origin of switchable quasiparticle-interference chirality in loop-current phase of kagome metals measured by scanning-tunneling-microscopy 
}%
\end{center}

\begin{center}
Seigo Nakazawa$^1$, Rina Tazai$^2$, Youichi Yamakawa$^1$, Seiichiro Onari$^1$, and Hiroshi Kontani$^1$
\end{center}

\begin{center}
$^1$ \textit{Department of Physics, Nagoya University, Nagoya 464-8602, Japan}

$^2$ \textit{Yukawa Institute for Theoretical Physics, Kyoto University,
Kyoto 606-8502, Japan}
\end{center}

\end{widetext}



\subsection{A: Tight-binding model with BO and LC order parameters}

\begin{figure}[htb]
\includegraphics[width=0.99\linewidth]{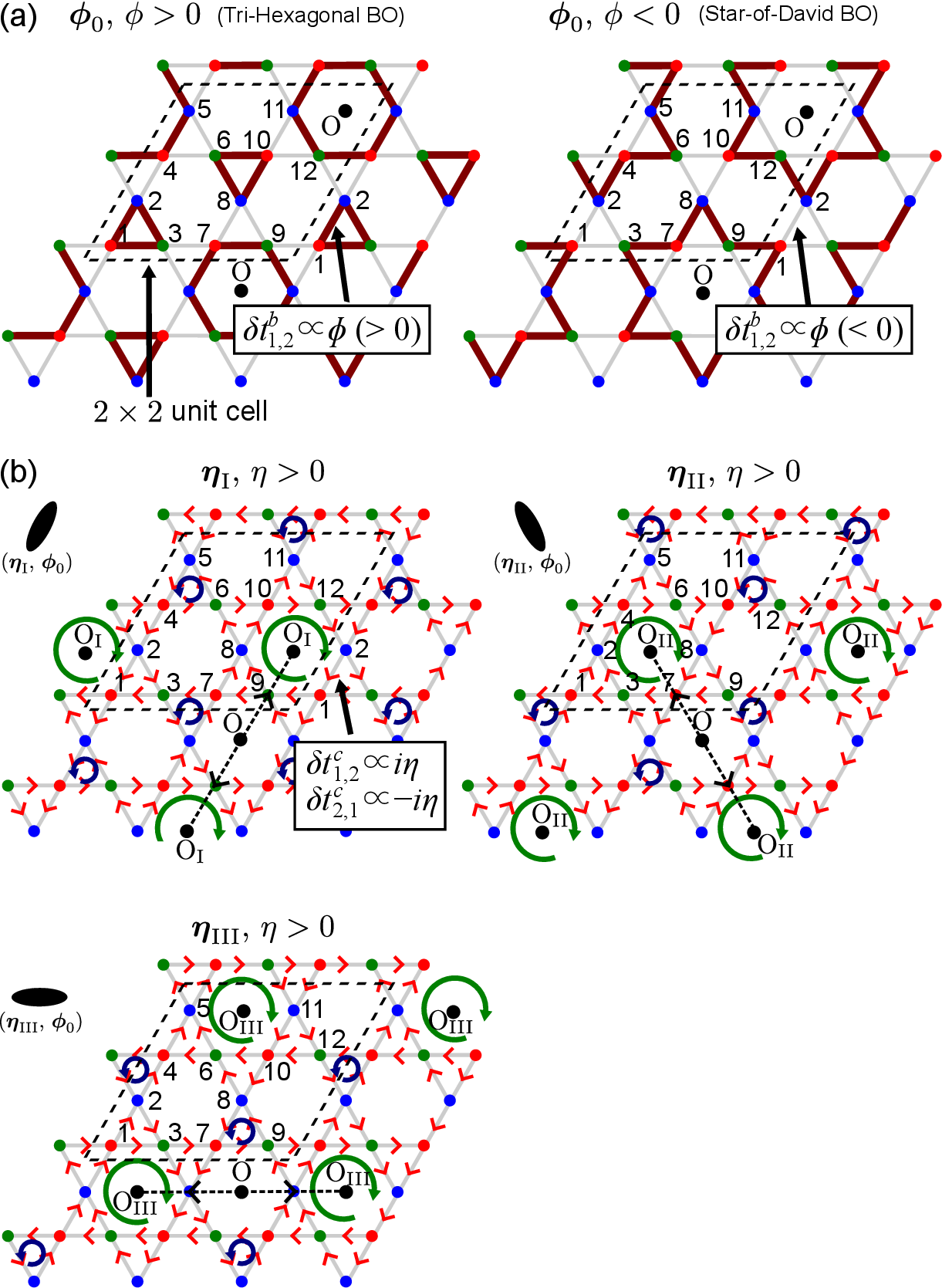}
\caption{
(a) BO parameter ${\bm\phi}_0$ in real space for TrH pattern ($\phi>0$) and SoD pattern ($\phi<0$).
(b) LC parameter ${\bm\eta}_{\a}$ ($\a$=I, II, III) in real space ($\eta>0$).
Sites A, B, and C in the main text (in Fig. 1 (b) and Fig. 3 (c)) respectively correspond to sites 10, 8, and 12.  
}
\label{fig:figS1}
\end{figure}

The bond/current order is the modulation of the hopping integral between $i$ and $j$ atoms due to the electron correlation, $\delta t_{ij}^{\rm b/c}$.
Theoretically, it is the symmetry breaking in the self-energy, and it is derived from the density-wave (DW) equation \cite{Tazai1S,Tazai2S}.
The wavevectors of the bond and current orders correspond to the inter-sublattice nesting vectors $\q_n$ ($n=1-3$).
The triple-$Q$ ($3Q$) bond order between the nearest V atoms is given as 
%
\begin{eqnarray}
\delta t_{ij}^{\rm b}&=& \phi_1 {g}_{ij}^{(1)}+\phi_2 {g}_{ij}^{(2)}+\phi_3 {g}_{ij}^{(3)} ,
\label{eqn:tb}
\end{eqnarray}
where ${\bm \phi}\equiv(\phi_1,\phi_2,\phi_3)$ is the set of BO parameters
with the wavevector $\q_n$,
and $g_{ij}^{(n)}=g_{ji}^{(n)}=\pm1$ for the nearest-neighbor sites $(i,j)$ is the even-parity form factor for the BO.
For $n=1$, $g_{ij}^{(1)}=+1 \ [-1]$ for 
sites $(i,j)$ belongs to sublattices $(1,2),(4,5),(8,10),(11,7)$ [$(2,4),(5,1),(7,8),(10,11)$].
Figure \ref{fig:figS1} (a) represents the $3Q$ bond order ${\bm\phi}=\phi{\bm e}_0$, where ${\bm e}_0=(1,1,1)/\sqrt{3}$.
The TrH (SoD) bond-order is realized for $\phi>0$ ($\phi<0$).
The enlarged unit cell contains twelve sublattices ($l=1\sim12$).
Sites A, B, and C in the main text respectively correspond to sites 10, 8, and 12.

Next, we explain the $3Q$ current order between the nearest V atoms.
Its form factor with ${\bm q}={\bm q}_1$, $f_{ij}^{(1)}$, is 
$+i$ for sites $(i,j)$ belongs to sublattices $(l,m)=(1,2),(2,4),(4,5),(5,1)$, and $-i$ for $(7,8),(8,10),(10,11),(11,7)$.
Odd parity relation $f_{ij}^{(1)}=-f_{ji}^{(1)}$ holds.
Other form factors with ${\bm q}_2$ and ${\bm q}_3$, 
$f_{ij}^{(2)}$ and $f_{ij}^{(3)}$, are also derived from Fig. \ref{fig:figS1} (b).
Using ${\bm f}_{ij}=(f_{ij}^{(1)},f_{ij}^{(2)},f_{ij}^{(3)})$, 
the current order is 
\begin{eqnarray}
\delta t_{ij}^{\rm c}&=& 
\eta_1 f_{ij}^{(1)}+\eta_2 f_{ij}^{(2)}+\eta_3 f_{ij}^{(3)} ,
\label{eqn:tc}
\end{eqnarray}
where ${\bm \eta}\equiv(\eta_1,\eta_2,\eta_3)$ is the set of 
current order parameters with the wavevector $\q_n$.
Figure \ref{fig:figS1} (a) represents the $3Q$ current order ${\bm\eta}=\eta{\bm e}_\a$, 
where ${\bm e}_{\rm I}=(a,-1,-1)/\sqrt{2+a^2}$, ${\bm e}_{\rm II}=(-1,a,-1)/\sqrt{2+a^2}$, and ${\bm e}_{\rm III}=(-1,-1,a)/\sqrt{2+a^2}$.
When $|\eta/\phi|\ll1$, 
the LC + BO coexisting state is given by 
(${\bm\eta}_\a$, ${\bm\phi}_0$) with $a\lesssim2$
according to the GL free-energy analysis.
(For simplicity, we set $a=2$ in this paper, although similar results are obtained for $a=1$.)
Then, the electronic states become nematic, with the director aligned the ab direction for $\a$=I, the bc direction for $\a$=II, and the ca direction for $\a$=III.
In the theoretical model of the main text,
the tight-binding hopping integral is given as
$t_{ij}= t_{ij}^0+\delta t_{ij}^{\rm b}+\delta t_{ij}^{\rm c}$
\cite{Nakazawa-impS,ShimuraS}.

\begin{figure}[htb]
\includegraphics[width=0.99\linewidth]{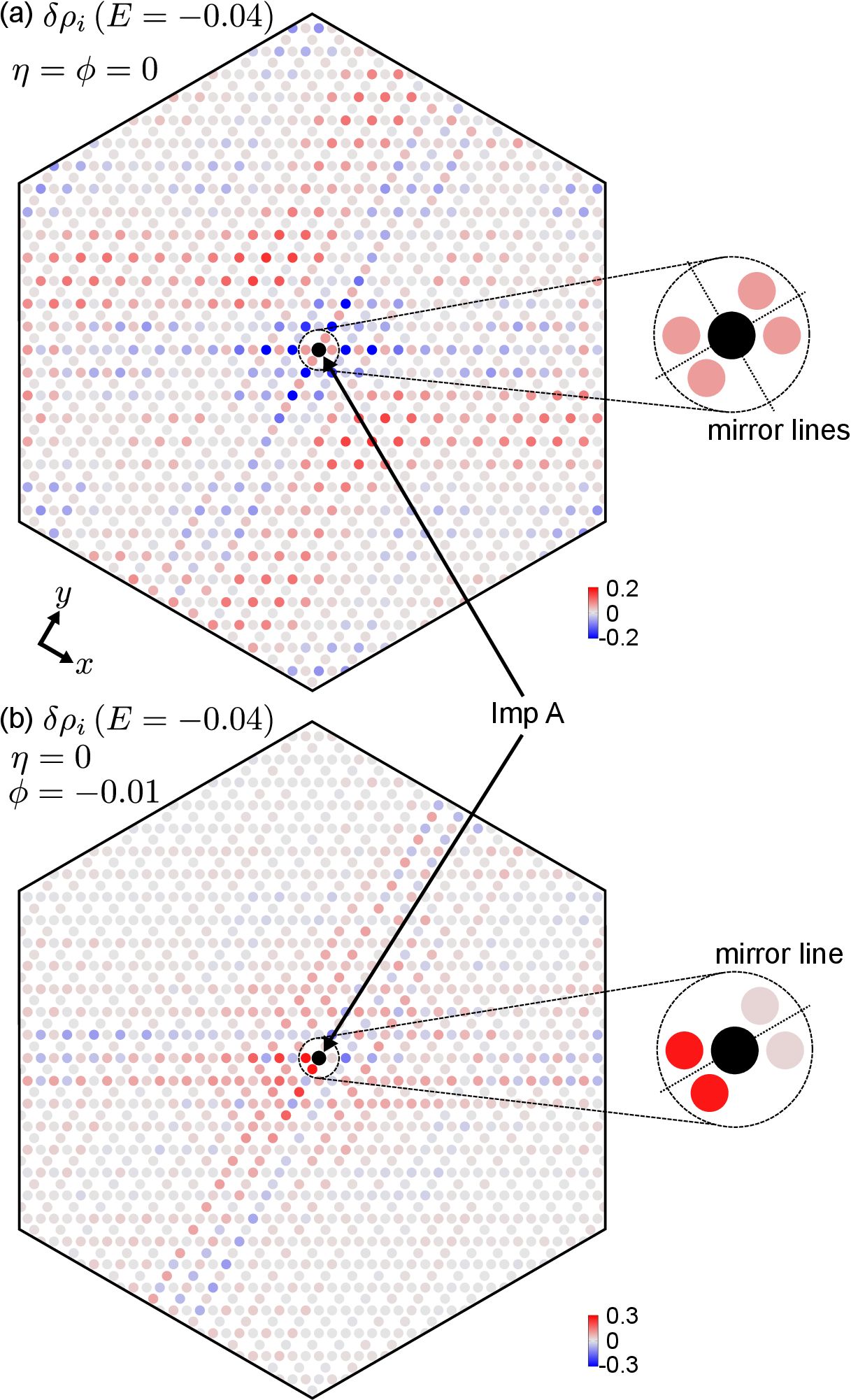}
\caption{
Modulation of the LDOS ($E=-0.04$) induced by single Imp A in $N=1452$-site unit cell model in the absence of the LC order, for (a) no BO ($\phi=0$) and (b) finite BO ($\phi=-0.01$).
Both LDOS patterns are achiral and exhibit $C_{1h}$ symmetry around Imp A.
}
\label{fig:figS1-2}
\end{figure}

Figure \ref{fig:figS1-2} shows the modulation of the LDOS ($E=-0.04$) induced by a single Imp A in the absence of the LC order, for (a) no BO ($\phi=0$) and (b) finite BO ($\phi=-0.01$).
In both cases, the LDOS pattern exhibits mirror symmetry with respect to the mirror lines shown as dotted lines in Fig. \ref{fig:figS1-2}.
For this reason, impurity-induced QPI signal is achiral when the LC order is absent ($\eta=0$).

In highly contrast, in the nematic LC + BO state, the LDOS pattern violates any mirror symmetries, as demonstrated in Fig. \ref{fig:fig2} (a) in the main text.
For this reason, chiral QPI signal can be caused by dilute impurities.
Notably, in the nematic LC + BO state, the chirality of this QPI signal can be flipped by applying small magnetic field $B_z$.
This is the main finding of this paper.

\subsection{B: GL free-energy for LC + BO state}

To discuss the change in the QPI chirality by Imp Z ($\chi_{\rm Z}=\pm1, 0$) under small magnetic field, we analyze the GL free energy with respect to the LC and BO order parameters.
The free-energy up to the fourth order terms is given as
\cite{Tazai2S,BalentsS}
$F[{\bm\eta},{\bm\phi_0}]= a_b |{\bm\phi}|^2 +a_c |{\bm\eta}|^2 + b_1 \phi_1\phi_2\phi_3 + b_2 (\phi_1\eta_2\eta_3+{\rm cycl.}) + c_1(\phi_1^4+\phi_2^4+\phi_3^4) + c_2(\phi_1^2\phi_2^2+{\rm cycl.})+ c_3(\eta_1^4+\eta_2^4+\eta_3^4) + c_4(\eta_1^2\eta_2^2+{\rm cycl.}) + 2c_5(\phi_1^2\eta_1^2+\phi_2^2\eta_2^2+\phi_3^2\eta_3^2)+ c_6(\phi_1^2(\eta_2^2+\eta_3^2)+{\rm cycl.})$,
where $a_b\propto(T_{\rm cdw}-T)$ and $a_c\propto(T_{\rm lc}-T)$ are the second-order,
$b_1$ and $b_2$ are the third-order,
and $c_1\sim c_6$ are the fourth-order parameters. 
Here, we assume ${\bm\phi}={\bm\phi}_0$.
Then, the $\phi^3$ term becomes $b_1 \phi^3/3\sqrt{3}$,
so the relation $b_1\phi<0$ is satisfied to stabilize the 3Q BO.
Then, the GL free energy is simplified as
$F[{\bm\eta},{\bm\phi_0}]= (a_c - b_1 \phi/2\sqrt{3} +2(c_1+c_2)\phi^2/3) |{\bm\eta}|^2 + (-b_1 \phi/2\sqrt{3})(\eta_1+\eta_2+\eta_3)^2 +c_1|{\bm\eta}|^4 + (c_2-2c_1)(\eta_1^2\eta_2^2+{\rm cycl.})$.
Here, we used the approximate relation $b_2=-b_1$, $c_1=c_3=c_5$, and $c_2=c_4=c_6$.
Because $b_1\phi<0$, the relation $\eta_1+\eta_2+\eta_3=0$ is satisfied for $|{\bm\phi}|\gg|{\bm\eta}|$.
The $3Q$ ($1Q$) LC order is realized when $c_2<2c_1$ ($c_2>2c_1$).
Thus, the stable $3Q$ LC + BO is ($\pm{\bm\eta}_{\a}$, ${\bm\phi}_0$),
where $\a$ = I, II, III.
Experimentally, the $1Q$ LC is observed above $T_{\rm cdw}$ in Ref. \cite{torqueS}, while it changes to the $3Q$ LC state in the CDW phase.
Therefore, the relations $|c_1|\gg(c_2-2c_1)$ is expected to be realized.
Based on the above considerations, 
the free-energy would be almost unchanged under the conditions
(i) $\eta_1^2+\eta_2^2+\eta_3^2\equiv \eta_0^2$ ($\eta_0$=constant) and (ii) $\eta_1+\eta_2+\eta_3=0$.
Thus, the free-energy is unchanged under the condition
$\eta_0^2= \eta_1^2+\eta_2^2+(\eta_1+\eta_2)^2=2\eta_1^2+2\eta_2^2+2\eta_1\eta_2$.

Next, we introduce the GL expansion of the orbital magnetization.
The orbital magnetization under the LC + BO state is
$M_{\rm orb}=m_1{\bm\phi}\cdot {\bm\eta} + m_2 (\eta_1\phi_2\phi_3+{\rm cycl.}) + m_3 \eta_1\eta_2\eta_3$
\cite{Tazai3S}.
When ${\bm\phi}={\bm\phi}_0$,
we obtain $M_{\rm orb}=-m_3/3\sqrt{6}$ for ${\bm\eta}=+{\bm\eta}_{\a}$,
and $M_{\rm orb}=+m_3/3\sqrt{6}$ for ${\bm\eta}=-{\bm\eta}_{\a}$.
Therefore, in the case of $m_3<0$,
the state ${\bm\eta}_{\a}$ is realized under $B_z>0$,
and it is changed to $-{\bm\eta}_{\a'}$ under $B_z<0$.

\subsection{C: QPI signals for Imp Z$''$ and Imp Z$'''$}

\begin{figure}[htb]
\includegraphics[width=0.99\linewidth]{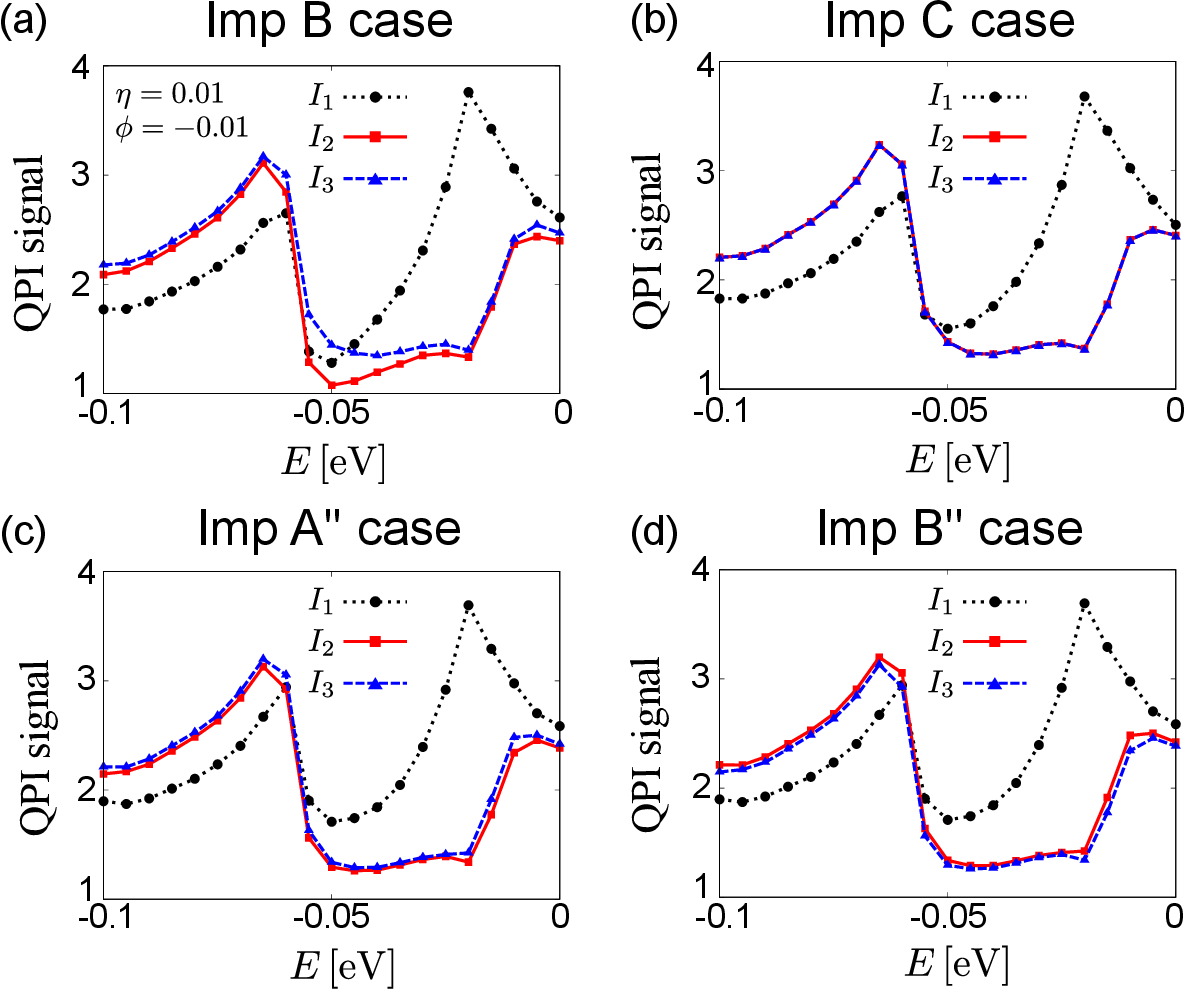}
\caption{
Obtained QPI signals for $({\bm\eta}_{\rm I},{\bm\phi}_0)$:
(a) In the case of Imp B, where the relation $I_2\ne I_3$ is realized.
Importantly, the obtained $I_2$ ($I_3$) is equivalent to $I_3$ ($I_2$) for Imp A, which is shown in the main text.
(b) In the case of Imp C, where the relation $I_2=I_3$ is realized.
(c) In the case of Imp ${\rm A}''$, where $I_2$ and $I_3$ are slightly different.
(d) In the case of Imp ${\rm B}''$, where $I_2$ ($I_3$) is equivalent to $I_3$ ($I_2$) for Imp ${\rm A}''$.
}
\label{fig:figS2}
\end{figure}

In Fig. 2 (d) in the main text, we demonstrate the $E$-dependence of the QPI signals $I_n$ ($n=1,2,3$) in the presence of the single impurity site A (=Imp A).
The same QPI signals are obtained for Imp A'.
The locations of sites A, A' and other sites in the 12 unit cell are shown in Fig. 1 (b).
Figure \ref{fig:figS2} (a) represents the QPI signals for Imp B.
(The same results are obtained for Imp B'.)
The obtained $I_2$ and $I_3$ are swapped with those in Fig. 2 (d) for Imp A.
Figure \ref{fig:figS2} (b) represents the QPI signals for Imp C,
where the relation $I_2=I_3$ is exactly satisfied.

We also demonstrate the QPI signals for Imp ${\rm A}''$ and Imp ${\rm B}''$ in Figs. \ref{fig:figS2} (c) and (d), respectively.
The locations of the sites ${\rm A}''$ and ${\rm B}''$ are shown in Fig. 1 (b).
In these cases, the relation $I_2 \ne I_3$ is realized as anticipated from the symmetry argument.
In addition, the obtained $I_2$ and $I_3$ for Imp ${\rm A}''$ are swapped with those for Imp ${\rm B}''$.
However, the obtained $I_2$ and $I_3$ are very close, so it is difficult to observe the QPI chirality for Imp ${\rm A}''$ and Imp ${\rm B}''$ cases.

\subsection{D: Chiral QPI signal due to Sb-site impurities}

\begin{figure}[htb]
\includegraphics[width=0.99\linewidth]{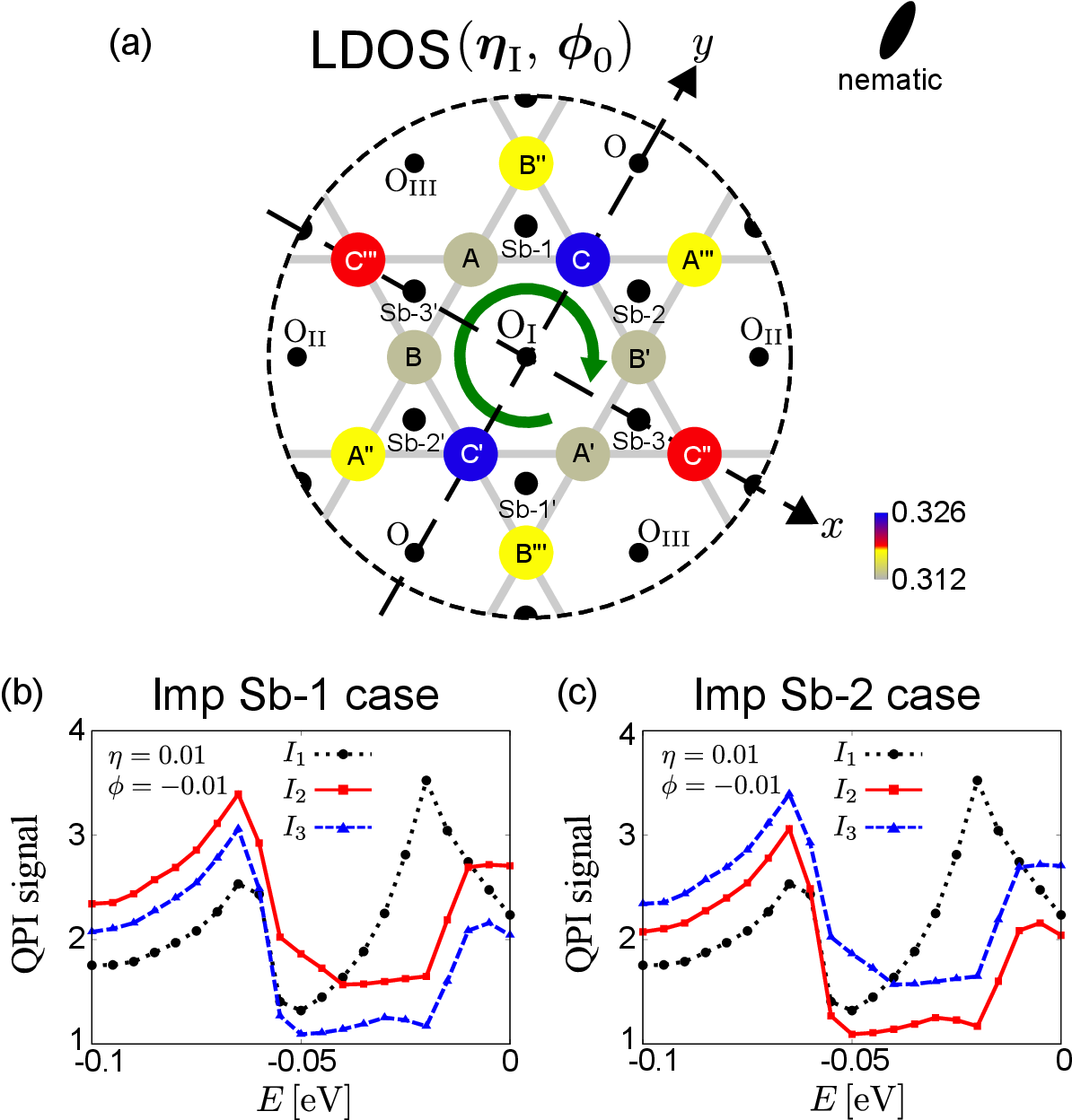}
\caption{
(a) Positions of Sb-site above the V-atom kagome lattice.
(b) Obtained chiral QPI signals for Sb-1 site impurity, (c) Sb-2 site impurity. Note that $I_2$ and $I_3$ in (b) are equal to $I_3$ and $I_2$ in (c), respectively.
}
\label{fig:figS4}
\end{figure}

In the main text, we discussed the QPI signal due to V-site impurity.
Here, we perform the same study for Sb-site impurities (or vacancy), whose locations are depicted in Fig. \ref{fig:figS4} (a).
They are above the V-atom kagome lattice plane.
We introduce the potential of the nearest V-sites due to the impurity at Sb-$n$ site, 
$V'(n_{\rm A}+n_{\rm B''}+n_{\rm C})$ for $n=1$,
$V'(n_{\rm A'''}+n_{\rm B'}+n_{\rm C})$ for $n=2$, and
$V'(n_{\rm A'}+n_{\rm B'}+n_{\rm C''})$ for $n=3$, and so on.
In the following numerical study, we set $V'=1$.

Figure \ref{fig:figS4} (b) represents the obtained QPI signal due to Sb-1 site 
single impurity for $N=1452$ ($M_x=M_y=11$),
for the LC + BO state $({\bm\eta}_{\rm I},{\bm\phi}_0)$ with $\eta=-\phi=0.01$.
Thus, we obtain the prominent chiral QPI signal, originating from the violation of all 2D mirror symmetries due to the Sb-1 site impurity.
The QPI chirality is reversed for Sb-2 site single impurity,
as shown in Fig. \ref{fig:figS4} (c).
In fact, the LDOS patterns with Sb-1 site impurity is converted to that with Sb-2 site through $M_x$ operation.
The same chirality is observed when a single impurity is introduced into Sb-1' and Sb-2', which are the inversions of Sb-1 and Sb-2, respectively.
In contrast, when a single impurity is introduced at other Sb sites, the QPI signal becomes nematic.


To summarize, prominent chiral QPI signal can be driven by impurities on Sb-sites under the nematic LC + BO phase in kagome metals.
The QPI chirality $\chi_{\rm Z'}=\pm1,0$ depends on the Sb-site on the impurity denoted as Z'. 
Interestingly, the QPI signal exhibits sizable chirality for wide range of $E$, compared with the V-site impurity case.
Importantly, the QPI chirality due to a fixed Sb-site impurity $\chi_{\rm Z}=\pm1,0$ is changed by reversing the magnetic field.

\subsection{E: Chiral QPI signal due to stripe CDW order in LC + BO phase}

\begin{figure}[htb]
\includegraphics[width=0.99\linewidth]{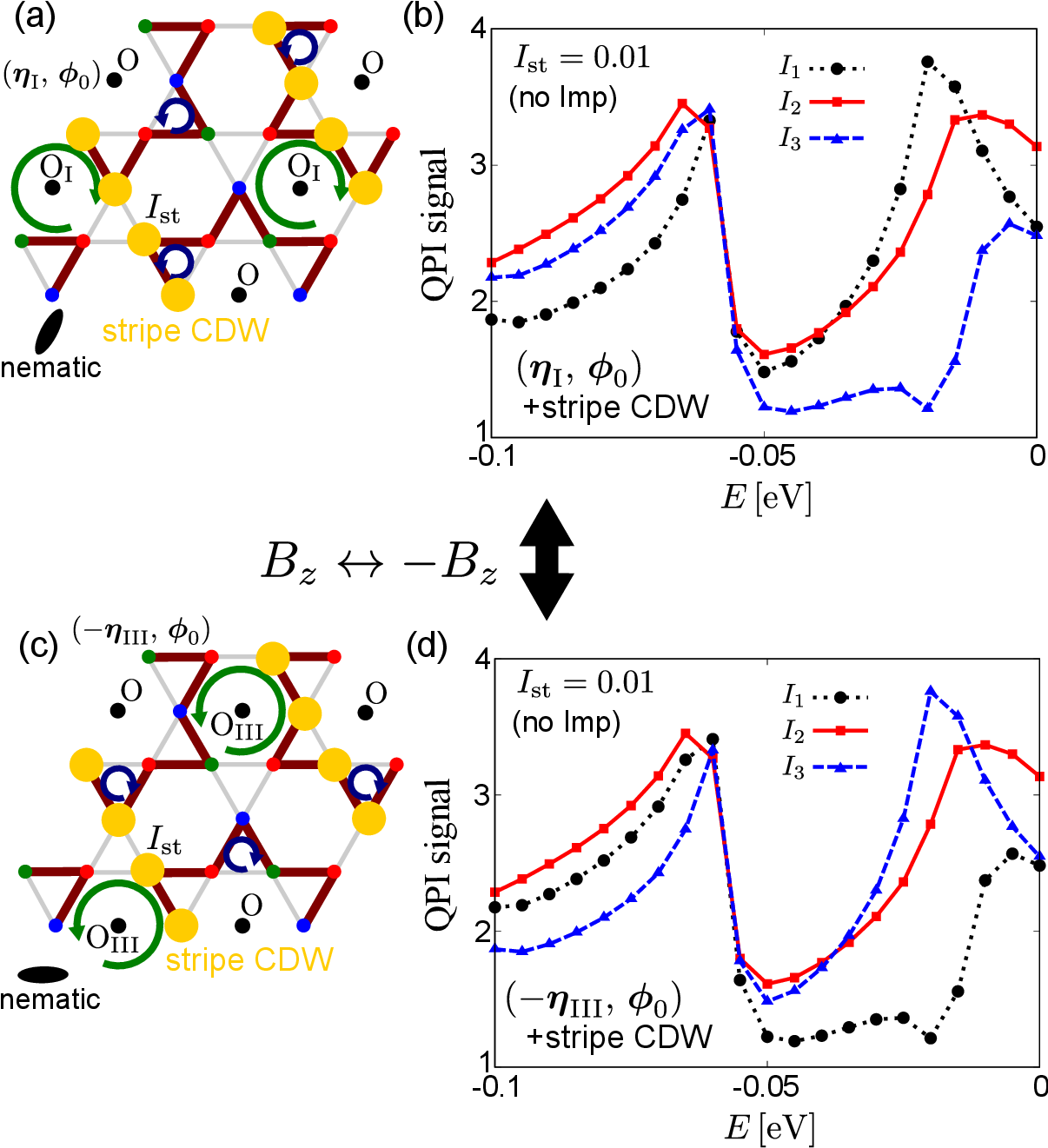}
\caption{
(a) LC + BO phase $({\bm\eta}_{\rm I},{\bm\phi}_0)$ with the $2a_0$ stripe CDW along ab direction, which is shown by the yellow circles.
(b) Obtained chiral QPI signals in the presence of the stripe CDW ($I_{\rm st}=0.01$) in the LC + BO phase.
The chirality is induced by the stripe CDW, even in the absence of the impurity. (c) LC + BO phase $(-{\bm\eta}_{\rm III},{\bm\phi}_0)$ with the $2a_0$ stripe CDW.
(d) Obtained chiral QPI signals, whose chirality is opposite to the chirality in (b).
}
\label{fig:figS3}
\end{figure}

Figure \ref{fig:figS3} (a) represents the LC + BO phase (${\bm\eta}_{\rm I}$, ${\bm\phi}_0$), whose nematic director is along ab axis.
Here, we also introduce the bc-direction stripe CDW with period $2a_0$.
The yellow circles represent the stripe CDW potentials $I_{\rm st}$,
which is consistent with the energy-independent (vertical) 
QPI signal with period $2a_0$ is universally observed in CsV$_3$Sb$_5$
 \cite{stripe1S,stripe2S}.
In Fig. \ref{fig:figS3} (a), the stripe CDW is along the bc axis.
When $I_{\rm st}=0$, the center of the hexagon with circulating current (indicated by the green arrow) is the center of inversion, and thus inversion symmetry exists.
However, the inversion symmetry is broken for $I_{\rm st}\ne0$ \cite{Tazai-eMChAS}. 
For this reason, the giant nonreciprocal transport phenomenon under the magnetic field, called the electronic magneto-chiral anisotropy (eMChA), is realized in kagome metals 
\cite{eMChAS,Tazai-eMChAS}. 
Figure \ref{fig:figS3} (b) represents the obtained QPI signals as functions of $E$, in the presence of the stripe CDW ($I_{\rm st}=0.01$) in the LC + BO phase.
When $I_{\rm st}=0$, the relation $I_1\ne I_2=I_3$ is realized.
However, $I_2$ and $I_3$ become inequivalent when $I_{\rm st}\ne0$.
Therefore, the chirality is induced by the stripe CDW, even in the absence of the impurity.

The nematic LC + BO state (${\bm\eta}_{\rm I}$, ${\bm\phi}_0$) will be flipped to  ($-{\bm\eta}_{\rm III}$, ${\bm\phi}_0$) or ($-{\bm\eta}_{\rm II}$, ${\bm\phi}_0$) \cite{Tazai-eMChAS}.
The state ($-{\bm\eta}_{\rm III}$, ${\bm\phi}_0$) is schematically expressed in Fig. \ref{fig:figS3} (c).
Figure \ref{fig:figS3} (d) shows the obtained QPI signal,
where the signals $I_1$ and $I_3$ in Fig. \ref{fig:figS3} (b) are exchanged.
(When $I_{\rm st}=0$, the relation $I_3\ne I_1=I_2$ is realized.)
Importantly, the QPI chirality in Fig. \ref{fig:figS3} (b) and that in Fig. \ref{fig:figS3} (d) are opposite:
For example, at $E=-0.1$, the relation $I_2>I_3>I_1$ holds in Fig. \ref{fig:figS3} (b), while the relation $I_2>I_1>I_3$ holds in Fig. \ref{fig:figS3} (d).
The obtained results are consistent with the experimental report for A=Rb in Ref. \cite{STM-MadS}.

\begin{thebibliography}{999}
\bibitem{Ortiz1}
    B. R. Ortiz, L. C. Gomes, J. R. Morey, M. Winiarski, M. Bordelon, J. S. Mangum, 
    I. W. H. Oswald, J. A. Rodriguez-Rivera, J. R. Neilson, S. D. Wilson, E. Ertekin, 
    T. M. McQueen, and E. S. Toberer, 
    New kagome prototype materials: discovery of KV$_3$Sb$_5$, RbV$_3$Sb$_5$, 
    and CsV$_3$Sb$_5$, 
    Phys. Rev. Materials \textbf{3}, 094407 (2019).
    
    \bibitem{Ortiz2}
    B. R. Ortiz, S. M. L. Teicher, Y. Hu, J. L. Zuo, P. M. Sarte, E. C. Schueller, 
    A. M. M. Abeykoon, M. J. Krogstad, S. Rosenkranz, R. Osborn, R. Seshadri, 
    L. Balents, J. He, and S. D. Wilson, 
    CsV$_3$Sb$_5$: A $\mathbb{Z}_2$ Topological kagome metal with a 
    superconducting ground state, 
    Phys. Rev. Lett. \textbf{125}, 247002 (2020).
    
    \bibitem{BO2}
    H. Li, H. Zhao, B. R. Ortiz, T. Park, M. Ye, L. Balents, Z. Wang, 
    S. D. Wilson, and I. Zeljkovic, 
    Rotation symmetry breaking in the normal state of 
    a kagome superconductor KV$_3$Sb$_5$, 
    Nat. Phys. \textbf{18}, 265 (2022).
    
    \bibitem{BO3}
    L. Nie, K. Sun, W. Ma, D. Song, L. Zheng, Z. Liang, P. Wu, F. Yu, J. Li, 
    M. Shan, D. Zhao, S. Li, B. Kang, Z. Wu, Y. Zhou, K. Liu, Z. Xiang, J. Ying, 
    Z. Wang, T. Wu, and X. Chen, 
    Charge-density-wave-driven electronic nematicity in a kagome superconductor, 
    Nature \textbf{604}, 59 (2022).
    
    \bibitem{BO4}
    T. Kato, Y. Li, T. Kawakami, M. Liu, K. Nakayama, Z. Wang, A. Moriya, K. Tanaka, 
    T. Takahashi, Y. Yao, and T. Sato, 
    Three-dimensional energy gap and origin of charge-density wave 
    in kagome superconductor KV$_3$Sb$_5$, 
    Commun. Mater. \textbf{3}, 30 (2022).
    
    \bibitem{nematicity1}
    L. Nie, K. Sun, W. Ma, D. Song, L. Zheng, Z. Liang, P. Wu, F. Yu, J. Li, M. Shan, 
    D. Zhao, S. Li, B. Kang, Z. Wu, Y. Zhou, K. Liu, Z. Xiang, J. Ying, Z. Wang, 
    T. Wu, and X. Chen, 
    Charge-density-wave-driven electronic nematicity in a kagome superconductor, 
    Nature \textbf{604}, 59 (2022).
    
    \bibitem{nematicity2}
    Y. Xu, Z. Ni, Y. Liu, B. R. Ortiz, Q. Deng, S. D. Wilson, B. Yan, L. Balents, 
    and L. Wu, 
    Three-state nematicity and magneto-optical Kerr effect in the charge density waves 
    in kagome superconductors, 
    Nat. Phys. \textbf{18}, 1470 (2022).
    
    \bibitem{torque}
    T. Asaba, A. Onishi, Y. Kageyama, T. Kiyosue, K. Ohtsuka, S. Suetsugu, 
    Y. Kohsaka, T. Gaggl, Y. Kasahara, H. Murayama, K. Hashimoto, R. Tazai, 
    H. Kontani, B. R. Ortiz, S. D. Wilson, Q. Li, H.-H. Wen, T. Shibauchi, 
    and Y. Matsuda, 
    Evidence for an odd-parity nematic phase above the charge-density-wave transition 
    in a kagome metal, 
    Nat. Phys. \textbf{20}, 40 (2024).

    \bibitem{Tazai1}
    R. Tazai, Y. Yamakawa, S. Onari, and H. Kontani, 
    Mechanism of exotic density-wave and beyond-Migdal unconventional superconductivity 
    in kagome metal AV$_3$Sb$_5$ (A=K, Rb, Cs), 
    Sci. Adv. \textbf{8}, eabl4108 (2022).
    
    \bibitem{SC1}
    M. Roppongi, K. Ishihara, Y. Tanaka, K. Ogawa, K. Okada, S. Liu, K. Mukasa, 
    Y. Mizukami, Y. Uwatoko, R. Grasset, M. Konczykowski, B. R. Ortiz, S. D. Wilson, 
    K. Hashimoto, and T. Shibauchi, 
    Bulk evidence of anisotropic $s$-wave pairing with no sign change 
    in the kagome superconductor CsV$_3$Sb$_5$, 
    Nat Commun \textbf{14}, 667 (2023).
    
    \bibitem{SC2}
    W. Zhang, X. Liu, L. Wang, C. W. Tsang, Z. Wang, S. T. Lam, W. Wang, J. Xie, 
    X. Zhou, Y. Zhao, S. Wang, J. Tallon, K. T. Lai, and S. K. Goh, 
    Nodeless superconductivity in kagome metal CsV$_3$Sb$_5$ with 
    and without time reversal symmetry breaking, 
    Nano Lett. \textbf{23}, 872 (2023).

    \bibitem{STM}
    Y.-X. Jiang, J.-X. Yin, M. M. Denner, N. Shumiya, B. R. Ortiz, G. Xu, Z. Guguchia, 
    J. He, M. S. Hossain, X. Liu, J. Ruff, L. Kautzsch, S. S. Zhang, G. Chang, 
    I. Belopolski, Q. Zhang, T. A. Cochran, D. Multer, M. Litskevich, Z.-J. Cheng, 
    X. P. Yang, Z. Wang, R. Thomale, T. Neupert, S. D. Wilson, and M. Z. Hasan, 
    Unconventional chiral charge order in kagome superconductor KV$_3$Sb$_5$, 
    Nat. Mater. \textbf{20}, 1353 (2021).

\bibitem{STM-Yin}
H. Deng {\it et al}.,
Chiral kagome superconductivity modulations with residual Fermi arcs, 
Nature {\bf 632}, 775 (2024).


\bibitem{STM-Mad}
Y. Xing, S. Bae, E. Ritz, Fan Yang, Turan Birol, Andrea N. Capa Salinas, Brenden R. Ortiz, Stephen D. Wilson, Z. Wang, R. M. Fernandes, and V. Madhavan,
Nature {\bf 631}, 60 (2024).

\bibitem{STM-PRB1}
Z. Wang {\it et al}., 
Electronic nature of chiral charge order in the kagome superconductor CsV$_3$Sb$_5$,
Phys. Rev. B {\bf 104}, 075148 (2021).

\bibitem{STM-PRB2}
N. Shumiya {\it et al}., 
Intrinsic nature of chiral charge order in the kagome superconductor RbV$_3$Sb$_5$,
Phys. Rev. B {\bf 104}, 035131 (2021).

\bibitem{Varma}
C. M. Varma, 
Non-Fermi-liquid states and pairing instability of a general model of copper oxide metals,
Phys. Rev. B \textbf{55}, 14554 (1997).

\bibitem{Varma2}
C. Weber, T. Giamarchi, and C. M. Varma,
{\it Phase Diagram of a Three-Orbital Model for High-$T_{\rm c}$ Cuprate Superconductors},
Phys. Rev. Lett. {\bf 112}, 117001 (2014).

    \bibitem{Balents}
    T. Park, M. Ye, and L. Balents, 
    Electronic instabilities of kagome metals: Saddle points and Landau theory, 
    Phys. Rev. B \textbf{104}, 035142 (2021).
    
    \bibitem{cLC}
    Y.-P. Lin and R. M. Nandkishore, 
    Complex charge density waves at Van Hove singularity on hexagonal lattices: 
    Haldane-model phase diagram and potential realization in the kagome metals 
    AV$_3$Sb$_5$ (A=K, Rb, Cs), 
    Phys. Rev. B 104, 045122 (2021).

    \bibitem{BO_theory1}
    M. L. Kiesel, C. Platt, and R. Thomale, 
    Unconventional Fermi surface instabilities in the kagome Hubbard model, 
    Phys. Rev. Lett. \textbf{110}, 126405 (2013).
    
    \bibitem{BO_theory2}
    W.-S. Wang, Z.-Z. Li, Y.-Y. Xiang, and Q.-H. Wang, 
    Competing electronic orders on kagome lattices at van Hove filling, 
    Phys. Rev. B \textbf{87}, 115135 (2013).
    
    \bibitem{BO_theory3}
    X. Wu, T. Schwemmer, T. Müller, A. Consiglio, G. Sangiovanni, D. D. Sante, 
    Y. Iqbal, W. Hanke, A. P. Schnyder, M. M. Denner, M. H. Fischer, T. Neupert, 
    and R. Thomale, 
    Nature of Unconventional Pairing in the Kagome Superconductors AV$_3$Sb$_5$ 
    (A = K, Rb, Cs), 
    Phys. Rev. Lett. \textbf{127}, 177001 (2021).
    
    \bibitem{BO_theory4}
    M. M. Denner, R. Thomale, and T. Neupert, 
    Analysis of Charge Order in the Kagome Metal AV$_3$Sb$_5$ (A = K, Rb, Cs), 
    Phys. Rev. Lett. 127, 217601 (2021).

    \bibitem{Fernandes}
    M. H. Christensen, T. Biro, B. M. Andersen, and R. M. Fernandes,
    Loop currents in AV$_3$Sb$_5$ kagome metals: Multipolar and toroidal magnetic order,
    Phys. Rev. B \textbf{106}, 144504 (2022).

    \bibitem{AHE1}
    S.-Y. Yang, Y. Wang, B. R. Ortiz, D. Liu, J. Gayles, E. Derunova, 
    R. Gonzalez-Hernandez, Sejkal, Y. Chen, S. S. P. Parkin, S. D. Wilson, 
    E. S. Toberer, T. McQueen, and M. N. Ali, 
    Giant, unconventional anomalous Hall effect 
    in the metallic frustrated magnet candidate, KV$_3$Sb$_5$, 
    Sci. Adv. \textbf{6}, eabb6003 (2020).
    
    \bibitem{AHE2}
    F. H. Yu, T. Wu, Z. Y. Wang, B. Lei, W. Z. Zhuo, J. J. Ying, and X. H. Chen, 
    Concurrence of anomalous Hall effect and charge density wave 
    in a superconducting topological kagome metal, 
    Phys. Rev. B \textbf{104}, L041103 (2021).
    
    \bibitem{AHE3}
    Y. Wang, Z. Chen, Y. Nie, Y. Zhang, Q. Niu, G. Zheng, X. Zhu, W. Ning, 
    and M. Tian, 
    Nontrivial Fermi surface topology and large anomalous Hall effect 
    in the kagome superconductor RbV$_3$Sb$_5$, 
    Phys. Rev. B \textbf{108}, 035117 (2023).

    \bibitem{eMChA}
    C. Guo, C. Putzke, C. Konyzheva, S. Konyzheva, X. Huang, M. Gutierrez-Amigo, 
    I. Errea, D. Chen, M. G. Vergniory, C. Felser, M. H. Fischer, T. Neupert, 
    and P. J. W. Moll, 
    Switchable chiral transport in charge-ordered kagome metal CsV$_3$Sb$_5$, 
    Nature \textbf{611}, 461 (2022). 

\bibitem{SM}
Supplementary Information

    \bibitem{Tazai2}
    R. Tazai, Y. Yamakawa, and H. Kontani, 
    Charge-loop current order and $Z_3$ nematicity mediated by bond order fluctuations 
    in kagome metals, 
    Nat. Commun. \textbf{14}, 7845 (2023).

    \bibitem{mSR1}
    L. Yu, C. Wang, Y. Zhang, M. Sander, S. Ni, Z. Lu, S. Ma, Z. Wang, Z. Zhao, 
    H. Chen, K. Jiang, Y. Zhang, H. Yang, F. Zhou, X. Dong, S. L. Johnson, 
    M. J. Graf, J. Hu, H.-J. Gao, and Z. Zhao, 
    Evidence of a hidden flux phase 
    in the topological kagome metal CsV$_3$Sb$_5$, 
    arXiv:2107.10714 (avalable at https://arxiv.org/abs/2107.10714).
    
    \bibitem{mSR2}
    C. Mielke III, D. Das, J.-X. Yin, H. Liu, R. Gupta, Y.-X. Jiang, M. Medarde, 
    X. Wu, H. C. Lei, J. Chang, P. Dai, Q. Si, H. Miao, R. Thomale, T. Neupert, 
    Y. Shi, R. Khasanov, M. Z. Hasan, H. Luetkens, and Z. Guguchia, 
    Time-reversal symmetry-breaking charge order in a kagome superconductor, 
    Nature \textbf{602}, 245 (2022).
    
    \bibitem{mSR3}
    R. Khasanov, D. Das, R. Gupta, C. Mielke III, M. Elender, Q. Yin, Z. Tu, 
    C. Gong, H. Lei, E. T. Ritz, R. M. Fernandes, T. Birol, Z. Guguchia, 
    and H. Luetkens, 
    Time-reversal symmetry broken by charge order in CsV$_3$Sb$_5$, 
    Phys. Rev. Res. \textbf{4}, 023244 (2022).
    
    \bibitem{mSR4}
    Z. Guguchia, C. Mielke III, D. Das, R. Gupta, J.-X. Yin, H. Liu, Q. Yin, 
    M. H. Christensen, Z. Tu, C. Gong, N. Shumiya, M. S. Hossain, 
    T. Gamsakhurdashvili, M. Elender, P. Dai, A. Amato, Y. Shi, H. C. Lei, 
    R. M. Fernandes, M. Z. Hasan, H. Luetkens, and R. Khasanov, 
    Tunable unconventional kagome superconductivity in charge ordered 
    RbV$_3$Sb$_5$ and KV$_3$Sb$_5$, 
    Nat. Commun. \textbf{14}, 153 (2023).
    
    \bibitem{NMR}
    J. Luo, Z. Zhao, Y. Z. Zhou, J. Yang, A. F. Fang, H. T. Yang, H. J. Gao, 
    R. Zhou, and G.-q. Zheng, 
    Possible star-of-David pattern charge density wave with additional modulation 
    in the kagome superconductor CsV$_3$Sb$_5$, 
    npj Quantum Mater. 7, 30 (2022).


    
    \bibitem{Kerr2}
    Y. Hu, S. Yamane, G. Mattoni, K. Yada, K. Obata, Y. Li, Y. Yao, Z. Wang, 
    J. Wang, C. Farhang, J. Xia, Y. Maeno, and S. Yonezawa, 
    Time-reversal symmetry breaking in charge density wave 
    of CsV$_3$Sb$_5$ detected by polar Kerr effect, 
    arXiv:2208.08036 (avalable at https://arxiv.org/abs/2208.08036).

    \bibitem{strain}
    C. Guo, G. Wagner, C. Putzke, D. Chen, K. Wang, L. Zhang, M. Gutierrez-Amigo, 
    I. Errea, M. G. Vergniory, C. Felser, M. H. Fischer, T. Neupert, and P. J. W. Moll, 
    Correlated order at the tipping point in the kagome metal CsV$_3$Sb$_5$, 
    Nature Physics \textbf{20}, 579 (2024).

    \bibitem{Tazai3}
    R. Tazai, Y. Yamakawa, and H. Kontani, 
    Drastic magnetic-field-induced chiral current order and emergent 
    current-bond-field interplay in kagome metals, 
    Proc. Natl. Acad. of Sci. (PNAS) \textbf{121}, e2303476121 (2024).

    \bibitem{STM-Fujita}
    K. Fujita, C. K. Kim, I. Lee, J. Lee, M. H. Hamidian, I. A. Firmo, S. Mukhopadhyay, H. Eisaki, S. Uchida, M. J. Lawler, E. -A. Kim, and J. C. Seamus Davis, 
Simultaneous transitions in cuprate momentum-space topology and electronic symmetry breaking,
    Science {\bf 344}, 612 (2014).

    \bibitem{STM-Hanaguri}
    C. J. Butler, Y. Kohsaka, Y. Yamakawa, M. S. Bahramy, S. Onari, H. Kontani, T. Hanaguri, and S. Shamoto,
    Correlation-driven electronic nematicity in the Dirac semimetal BaNiS$_2$,
    Proc. Natl. Acad. Sci. USA, {\bf 119}, e2212730119 (2022).

    \bibitem{STM-PDW}
    S. Wang, P. Choubey, Y. X. Chong, W. Chen, W. Ren, H. Eisaki, S. Uchida, P. J.  Hirschfeld, and J. C. Seamus Davis,
    Scattering interference signature of a pair density wave state in the cuprate pseudogap phase,
    Nat. Commun. {\bf 12}, 6087 (2021).

    \bibitem{Thomale}
    F. Grandi, A. Consiglio, M. A. Sentef, R. Thomale, and D. M. Kennes, 
    Theory of nematic charge orders in kagome metals, 
    Phys. Rev. B \textbf{107}, 155131 (2023).

    \bibitem{Sushkov}
    H. D. Scammell, J. Ingham, T. Li, and O. P. Sushkov,
    Chiral excitonic order from twofold van Hove singularities in kagome metals,
    Nat. Commun. \textbf{14}, 605 (2023).

    \bibitem{Shimura}
    K. Shimura, R. Tazai, Y. Yamakawa, S. Onari, and H. Kontani, 
    Real-space loop current pattern in time-reversal-symmetry breaking phase 
    in kagome metals, 
    J. Phys. Soc. Jpn. \textbf{93}, 033704 (2024).

    \bibitem{Onari}
    S. Onari, Y. Yamakawa, and H. Kontani, 
    Sign-reversing orbital polarization in the nematic phase of FeSe due to the $C_2$ 
    symmetry breaking in the self-energy, 
    Phys. Rev. Lett. \textbf{116}, 227001 (2016).
    
    \bibitem{fRG}
    R. Tazai, Y. Yamakawa, and H. Kontani, 
    Emergence of charge loop current in the geometrically frustrated Hubbard model: 
    A functional renormalization group study, 
    Phys. Rev. B \textbf{103}, L161112 (2021).
    
    \bibitem{DW_equation}
    R. Tazai, S. Matsubara, Y. Yamakawa, S. Onari, and H. Kontani, 
    Rigorous formalism for unconventional symmetry breaking in Fermi liquid theory 
    and its application to nematicity in FeSe, 
    Phys. Rev. B \textbf{107}, 035137 (2023).
    
    \bibitem{Jianxin}
    J. Huang, R. Tazai, Y. Yamakawa, S. Onari, and H. Kontani, 
    Low temperature phase transitions inside the CDW phase in the kagome metals 
    AV$_3$Sb$_5$ (A=Cs,Rb,K): Significance of mixed-type Fermi surface 
    electron correlations, 
    Phys. Rev. B \textbf{109}, L041110 (2024).

\bibitem{Kontani-rev}
H. Kontani, R. Tazai, Y. Yamakawa, and S. Onari,
{\it Unconventional density waves and superconductivities in Fe-based superconductors and other strongly correlated electron systems},
Adv. Phys. {\bf 70}, 355 (2021).

\bibitem{Nakazawa-imp}
S. Nakazawa, R. Tazai, Y. Yamakawa, S. Onari, and H. Kontani,
Giant Impurity Effects on Charge Loop Current Order States in Kagome Metals,
Phys. Rev. B {\bf 111}, 075161 (2025).

\bibitem{stripe1}
H. Zhao {\it et al.}, 
{\it Cascade of correlated electron states in the kagome superconductor CsV$_3$Sb$_5$},
Nature {\bf 599}, 216 (2021).

\bibitem{stripe2}
H. Li, H. Zhao, B. R. Ortiz, Y. Oey, Z. Wang, S. D. Wilson, and I. Zeljkovic,
{\it Unidirectional coherent quasiparticles in the high-temperature rotational symmetry broken phase of AV$_3$Sb$_5$ kagome superconductors},
Nat. Phys. {\bf 19}, 637 (2023).

\bibitem{Tazai-eMChA}
R. Tazai, Y. Yamakawa, T. Morimoto, and H. Kontani,
Quantum-metric-induced giant and reversible nonreciprocal transport phenomena in chiral loop-current phases of kagome metals,
arXiv:2408.04233.


    
\end{thebibliography}

\begin{thebibliography}{999}

    \bibitem{Tazai1S}
    R. Tazai, Y. Yamakawa, S. Onari, and H. Kontani, 
    Mechanism of exotic density-wave and beyond-Migdal unconventional superconductivity 
    in kagome metal AV$_3$Sb$_5$ (A=K, Rb, Cs), 
    Sci. Adv. \textbf{8}, eabl4108 (2022).
    
    \bibitem{Tazai2S}
    R. Tazai, Y. Yamakawa, and H. Kontani, 
    Charge-loop current order and $Z_3$ nematicity mediated by bond order fluctuations 
    in kagome metals, 
    Nat. Commun. \textbf{14}, 7845 (2023).

    \bibitem{ShimuraS}
    K. Shimura, R. Tazai, Y. Yamakawa, S. Onari, and H. Kontani, 
    Real-space loop current pattern in time-reversal-symmetry breaking phase 
    in kagome metals, 
    J. Phys. Soc. Jpn. \textbf{93}, 033704 (2024).

    \bibitem{Nakazawa-impS}
    S. Nakazawa, R. Tazai, Y. Yamakawa, S. Onari, and H. Kontani,
    Giant Impurity Effects on Charge Loop Current Order States in Kagome Metals,
    arXiv:2405.12141.

    \bibitem{BalentsS}
    T. Park, M. Ye, and L. Balents, 
    Electronic instabilities of kagome metals: Saddle points and Landau theory, 
    Phys. Rev. B \textbf{104}, 035142 (2021).
    
    \bibitem{torqueS}
    T. Asaba, A. Onishi, Y. Kageyama, T. Kiyosue, K. Ohtsuka, S. Suetsugu, 
    Y. Kohsaka, T. Gaggl, Y. Kasahara, H. Murayama, K. Hashimoto, R. Tazai, 
    H. Kontani, B. R. Ortiz, S. D. Wilson, Q. Li, H.-H. Wen, T. Shibauchi, 
    and Y. Matsuda, 
    Evidence for an odd-parity nematic phase above the charge-density-wave transition 
    in a kagome metal, 
    Nat. Phys. \textbf{20}, 40 (2024).

    \bibitem{Tazai3S}
    R. Tazai, Y. Yamakawa, and H. Kontani, 
    Drastic magnetic-field-induced chiral current order and emergent 
    current-bond-field interplay in kagome metals, 
    Proc. Natl. Acad. of Sci. (PNAS) \textbf{121}, e2303476121 (2024).

\bibitem{stripe1S}
H. Zhao {\it et al.}, 
{\it Cascade of correlated electron states in the kagome superconductor CsV$_3$Sb$_5$},
Nature {\bf 599}, 216 (2021).

\bibitem{stripe2S}
H. Li, H. Zhao, B. R. Ortiz, Y. Oey, Z. Wang, S. D. Wilson, and I. Zeljkovic,
{\it Unidirectional coherent quasiparticles in the high-temperature rotational symmetry broken phase of AV$_3$Sb$_5$ kagome superconductors},
Nat. Phys. {\bf 19}, 637 (2023).

\bibitem{Tazai-eMChAS}
R. Tazai, Y. Yamakawa, T. Morimoto, and H. Kontani,
Quantum-metric-induced giant and reversible nonreciprocal transport phenomena in chiral loop-current phases of kagome metals,
arXiv:2408.04233.

    \bibitem{eMChAS}
    C. Guo, C. Putzke, C. Konyzheva, S. Konyzheva, X. Huang, M. Gutierrez-Amigo, 
    I. Errea, D. Chen, M. G. Vergniory, C. Felser, M. H. Fischer, T. Neupert, 
    and P. J. W. Moll, 
    Switchable chiral transport in charge-ordered kagome metal CsV$_3$Sb$_5$, 
    Nature \textbf{611}, 461 (2022).

\bibitem{STM-MadS}
Y. Xing, S. Bae, E. Ritz, Fan Yang, Turan Birol, Andrea N. Capa Salinas, Brenden R. Ortiz, Stephen D. Wilson, Z. Wang, R. M. Fernandes, and V. Madhavan,
Nature {\bf 631}, 60 (2024).

\end{thebibliography}
\end{document}